\documentstyle[11pt,aasms4,psfig,epsf]{article}
\def\kms{km$\,{\rm s}^{-1}$}
\def\etal{\it et~al.\rm}
\def\sb{ergs cm$^{-2}$ s$^{-1}$ sr$^{-1}$}

\tighten
\begin{document}

\title{
Balmer-Dominated Spectra of Nonradiative Shocks in the Cygnus Loop,
RCW 86 and Tycho Supernova Remnants
}  

\vskip 1.0truein

\author{Parviz Ghavamian\altaffilmark{1,3,5}, John Raymond\altaffilmark{2}, 
R. Chris Smith\altaffilmark{4} and Patrick Hartigan\altaffilmark{1}}
\slugcomment{\it Accepted by the Astrophysical Journal}

\begin{abstract}

We present an observational and theoretical study of the optical emission from nonradiative
shocks in three supernova remnants: the Cygnus Loop, RCW 86 and Tycho.  The spectra of
these shocks are dominated by collisionally excited hydrogen Balmer lines which have both a broad component
caused by proton-neutral charge exchange and a narrow component caused by excitation of cold neutrals
entering the shock.  In each remnant we have obtained the broad to narrow flux ratios of the
H$\alpha$ and H$\beta$ lines and measured the H$\alpha$ broad component width.

A new numerical shock code computes the broad and narrow Balmer line emission
from nonradiative shocks in partially neutral gas.
The Balmer line fluxes are sensitive to Lyman line trapping and the degree of electron-proton
temperature equilibration.  The code calculates the density, temperature and size of the postshock
ionization layer and uses a Monte
Carlo simulation to compute narrow Balmer line enhancement from Lyman line trapping.  The initial fraction of the shock energy
allocated to the electrons and protons (the equilibration) is a free parameter.
Our models show that variations in electron-proton temperature equilibration and Lyman line trapping can reproduce the
observed range of broad to narrow ratios. 
The results give 80\%$-$100\% equilibration in nonradiative portions of the NE Cygnus Loop ($v_{S}\,\sim\,$300 \kms),
40\%$-$50\% equilibration in nonradiative portions of RCW 86 ($v_{S}\,\sim\,$600 \kms) and $\lesssim\,$20\%
equilibration in Tycho ($v_{S}\,\sim\,$2000 \kms).  Our results suggest an
inverse correlation between magnetosonic Mach number and equilibration in the observed remnants.

\keywords{ ISM: supernova remnants: shock waves, radiative transfer}
\end{abstract}
\altaffiltext{1}{Department of Space Physics and Astronomy, Rice University, 6100 S. Main St., 
Houston, TX, 77005$-$1892; hartigan@sparky.rice.edu}
\altaffiltext{2}{Harvard-Smithsonian Center for Astrophysics, 60 Garden St., Cambridge, 
MA 02138; raymond@cfa.harvard.edu}
\altaffiltext{3}{Current address: Department of Physics and Astronomy, Rutgers University, 136 Frelinghuysen Road, Piscataway,
NJ 08854-8019; parviz@physics.rutgers.edu}
\altaffiltext{4}{Cerro Tololo Inter-American Observatory, Casilla 603, Chile; csmith@noao.edu}
\altaffiltext{5}{Visiting Astronomer, Cerro Tololo Inter-American Observatory, National Optical Astronomy
Observatories.  CTIO is operated by AURA, Inc.\ under contract to the National Science
Foundation.}

\section{INTRODUCTION}

Supernova explosions produce some of the strongest shocks in nature.  During the event,
the outer layers of a star are ejected at speeds as high as 30,000 \kms.
The dense ejecta behave like a highly supersonic
piston, producing a strong shock wave in the ISM ahead of the piston (commonly known as the forward shock).
When the mass swept up by the forward shock begins to exceed the ejecta mass, the supernova remnant (SNR) enters
the blast wave (Sedov-Taylor) phase of evolution (Hamilton \& Sarazin 1984, Truelove \& McKee 1999).  
From the genesis of the SNR through the blast wave stage the forward shock is
nonradiative, meaning that it loses a negligible fraction of its internal energy to radiative cooling.

Due to the low density (n\,$\lesssim\,$1 cm$^{-3}$) and high Mach number ($M\,\gtrsim\,$200) of
the forward shock, the heated interstellar gas behaves like a collisionless plasma (Draine \& McKee 1993).
According to the strong shock jump conditions, electrons and protons are heated to temperatures in a minimum
ratio $T_{e}/T_{p}\,=\,m_{e}/m_{p}$\,($\sim\,$1/2000).  However, plasma waves and MHD turbulence at the shock
front may transfer energy from protons to electrons (Tidman \& Krall 1971, Kennel 1985, Cargill \& Papadapoulos 1988),
further equilibrating the two particle temperatures. 
Since the electrons and protons are affected by a different range of plasma waves and
the Coulomb equilibration time downstream often exceeds the age of the remnant, the two particle
temperatures can remain unequal throughout the shock. The types of plasma waves excited at the shock transition 
(and hence the amount of collisionless heating) depend strongly on parameters such as the magnetosonic Mach number 
$M_{S}$ and magnetic field orientation.  Due to this complicated dependence and the extreme difficulty of creating high
Mach number collisionless shocks in the laboratory, the properties of collisionless shocks remain poorly 
understood.

In a cold ($\lesssim\,$10$^{4}$ K), partially ionized medium, the charged and neutral distributions respond 
very differently to the passage of a collisionless shock.  The cold neutrals are initially unaffected by the
shock transition, while charged particles are compressed by a factor of four and strongly heated.  Some of the
cold neutrals entering the shock are
collisionally excited before being destroyed by collisional ionization or charge transfer.  Radiative decay 
by these excited neutrals produces narrow component Balmer
emission with a line width given by the preshock temperature.  In contrast, charge exchange between cold
neutrals and protons produces fast neutrals having the velocity distribution of the postshock protons.  Collisional 
excitation of the fast neutrals produces broad
Balmer line emission (Chevalier \& Raymond 1978, Chevalier, Kirshner \& Raymond 1980 (hereafter CKR80)). 
While the low temperatures in radiative shocks favor the emission of strong forbidden lines such as [O~II],
[O~III], [N~II] and [S~II], the high temperatures behind nonradiative shocks produce hydrogen Balmer line emission
far more efficiently.   Balmer-dominated shocks have been detected
in Tycho (Kirshner, Winkler \& Chevalier 1987 (hereafter KWC87), Smith \etal\, 1991,
Ghavamian \etal\, 2000), SN 1006 (KWC87, Smith \etal\, 1991, Winkler \& Long 1997),
RCW 86 (Long \& Blair 1990, Smith 1997), Kepler (Blair, Long \& Vancura 1991), portions of the Cygnus Loop
(Raymond \etal\, 1983 (hereafter RBFG83), Fesen \& Itoh 1985, Hester, Raymond \& Blair 1994 (hereafter HRB94)), four remnants in the LMC
(Tuohy \etal\, 1982, Smith \etal\, 1991, 1994) and the bow shock surrounding the pulsar PSR 1957+20
(Aldcroft, Romani \& Cordes 1992).

The optically emitting layer behind a nonradiative shock is extremely thin ($\lesssim\,$10$^{-3}$ pc), so
the proton temperature very close to the shock front can be measured from the width of a broad Balmer line.
The proton temperature in turn depends on the shock velocity through the jump conditions, making the broad line profile
a powerful diagnostic for probing the global kinematics of young SNRs.  The broad to narrow flux
ratio is another important diagnostic because it also depends on the shock velocity (CKR80, 
Smith \etal\, 1991).  Proper motion measurements 
of Balmer-dominated filaments have been combined with shock velocity estimates from the broad component to estimate
distances to several young SNRs (for example see Long, Blair \& van den Bergh 1988 for SN 1006, and Hesser \& 
van den Bergh 1981 for Tycho).

Balmer-dominated spectra can be difficult to interpret because the broad component
width and broad to narrow ratio yield different shock velocities depending on the assumed 
electron-proton equilibration.  Since the equilibration is not known a priori, spectroscopic observations in the past have
only yielded a range of shock velocities $v_{S}$ from the Balmer line profiles: a minimum $v_{S}(min)$ for no equilibration
and a maximum $v_{S}(max)$ for full equilibration.  In addition, collisional excitation behind the shock generates 
both Lyman line photons and
Balmer line photons.  Lyman line trapping by slow neutrals partially converts Ly $\beta$ and higher Lyman photons into 
Balmer line photons;
this enhances the Balmer line flux in the narrow component and complicates the diagnostic interpretation of the broad to narrow 
ratio (CKR80, Smith \etal\, 1991, Ghavamian 1999)).  Moreover, since the Ly $\beta$ optical depth is
larger than that of Ly $\gamma$, the H$\alpha$ broad to narrow ratio is more strongly affected than the H$\beta$ broad to 
narrow ratio.  The Lyman line optical depth behind the shock typically lies between 0 and 1, so that neither Case A
nor Case B conditions apply.  It is clear that disentangling the combined effects of Lyman line trapping and equilibration 
on a Balmer-dominated spectrum requires (1) the acquisition of high S/N line profiles in both H$\alpha$ and H$\beta$, 
and (2) careful modeling of the atomic physics and radiative transfer.

In \S 2 of this paper, we present
high S/N spectra of three nonradiative shocks which resolve the broad and narrow components of both
H$\alpha$ and H$\beta$.  In \S 3, we present measurements of the H$\alpha$ broad component width and broad to
narrow ratios in H$\alpha$ and H$\beta$.  We describe numerical models of nonradiative
shocks in \S 4, including calculations of ionization structures for a range of equilibrations
and Monte Carlo simulations of Ly $\beta$ and Ly $\gamma$ trapping.  In \S 5 we compare the observed broad to narrow ratios with
model predictions and simultaneously determine both the shock velocity and equilibration for the
observed SNRs.  In \S 6 we discuss our results and in \S 7 we present our conclusions.

\section{SPECTROSCOPIC OBSERVATIONS}

The spectroscopic datasets presented here
were acquired over a one year period between October 1997 and September 1998. 
We chose the instrumental setup for each observation to simultaneously maximize the number of detected photons
and provide enough spectral resolution to separate the broad Balmer lines from the narrow Balmer lines.
A log of our spectroscopic observations appears in Table~1.  The telescope and spectrograph configurations used
for the observations are described below.

\subsection{Cygnus Loop and Tycho}

Our spectroscopic observations of NE Cygnus and Tycho were performed
on the nights of October 7-10, 1997 (UT), using the 1.5 meter Tillinghast reflector telescope at the Fred
Lawrence Whipple Observatory.
The Schmidt camera CCD was a Loral 512$\times$2688 pixel chip, with an unbinned plate scale
of 0.6$^{\arcsec}$ per 15$\mu$m pixel.  To reduce readout noise during these observations, we binned the CCD chip 
by 4 pixels along the spatial direction.  The spectrograph was equipped with a 1200 l mm$^{-1}$ grating
blazed at 5700 \AA, providing a dispersion of 0.38 \AA\, pixel$^{-1}$ and spectral coverage
of 1000 \AA.  With this setup, only
one Balmer line could be fit on the CCD chip, so we performed the H$\alpha$ and H$\beta$ observations at
different times. Our target in NE Cygnus
Loop was filament P7 from the list of positions observed by the Hopkins Ultraviolet Telescope during the Astro-2 mission.
We used a 1.1$^{\arcsec}\times$3$^{\arcmin}$ slit for H$\alpha$, yielding a resolution of
1.4 \AA.  To obtain the H$\beta$ line profile, we observed Cygnus P7 again on the night of September 26, 1998 (UT).
In this case, we used a 3$^{\arcsec}\times$3$^{\arcmin}$ slit with the 1200 l mm$^{-1}$ grating, yielding a resolution of 1.5 \AA.
A POSS image of the Cygnus P7 filament appears in Figure 1, along with the two-dimensional FAST spectrum.   
The one-dimensional H$\alpha$ and H$\beta$ profiles appear in Figure 2.  From the H$\alpha$ line, we find a
broad component width of 262$\pm$32 \kms, with no evidence of forbidden line emission.  The shift between
the broad and narror component line centers is less than the uncertainty in broad component width, indicating
a nearly edge-on viewing geometry.

The 1997 Tycho observations focused on the bright nonradiative shock Knot g.  
In these observations, we used the 300 l mm$^{-1}$ grating with a 3$^{\arcsec}$ slit, 
centered on 5545 \AA\, (4000 \AA\, coverage). 
This convenient combination put both H$\alpha$ and H$\beta$ on the CCD chip and allowed simultaneous detection of
the H$\alpha$ and H$\beta$ broad components.  As in the previous observations, we binned the  CCD chip by 4 pixels.  The
resolution for this setup was 275 \kms\, (6 \AA\, at H$\alpha$).  It is evident from the two-dimensional spectrum
(Figure 3) that the H$\alpha$ surface brightness and broad component width vary with position along the Knot g filament,
reflecting variations in preshock density, shock velocity and viewing angle.  
The extraction aperture shown in Figure 3 was centered on the upper part of Knot g, where the broad component is brightest and
its width most nearly constant (hereafter  labeled Knot g (1)).  The extracted H$\alpha$ and H$\beta$
line profiles appear in Figure 4.  Knot g (1) is the brightest portion
of the filament (broad component FWHM of 1765$\pm$110 \kms), while the lower part Knot g (2) is fainter
(broad FWHM of 2105$\pm$130 \kms). The broad component center of Knot g (1) is redshifted 132$\pm$35 \kms\,
from the center of the narrow component, indicating that the plane of the shock is tilted
slightly into the plane of the sky. 

In the 1997 Tycho spectra, we detected faint diffuse H$\alpha$ emission above Knot g.  We did not subtract the
1-D diffuse spectrum directly from that of Knot g (1) because then the resulting 1-D spectrum becomes too noisy
to detect the H$\beta$ broad component.  Instead, we measured the H$\alpha$ surface brightnesses of Knot g and the
diffuse emission separately, using FAST spectra acquired in 1998 (using the same detector setup as the 1997
observations, see Ghavamian \etal\, 2000).  From the 1998 data, we found that the Diffuse H$\alpha$ 
emission contributes approximately 12\% of the narrow flux from Knot g (1).  We have used this information 
to correct the H$\alpha$ broad to narrow ratio of Knot g (1).  The diffuse H$\beta$ emission 
is very faint and is almost lost in the noise, so we have not corrected the H$\beta$ 
broad to narrow ratio.  In a separate paper (Ghavamian \etal\, 2000), we show that the diffuse emission is associated
with a photoionization precursor from the supernova blast wave.
Since the width of the broad Hbeta line and its shift from zero velocity must be the same as from the high S/N 
H$\alpha$ measurement, we hold
these quantities fixed while simultaneously fitting the broad and narrow H$\beta$ lines.  Constraining the fit in
this manner is useful because the S/N of the H$\beta$ broad component is signifcantly lower than that of H$\alpha$.

Prior to our observations, the most recent spectroscopic study of Knot g was that of Smith \etal\, (1991),
who found a broad FWHM of 1900$\pm$300 in H$\alpha$.  This value was obtained by averaging the broad component width over
the length of Knot g.  We note that this velocity lies roughly midway between our separately measured broad component
widths for sections (1) and (2).  In addition, the average H$\alpha$ broad to narrow ratio of sections (1) and (2)
lies within the range quoted by Smith \etal\, (1991).  However, in our spectra we find a significantly smaller shift
in the broad component line center than Smith \etal\, (1991).  
This may be due to the strong variations in shock viewing angle along the length of the filament.
In Table~2 we present our observational results for Knot g side by side with those of earlier papers.

Since the H$\alpha$ and H$\beta$ lines were recorded simultaneously, we were able to measure the broad and narrow component Balmer
decrements $I_{H\alpha}/I_{H\beta}$ separately for Knot g (1).  Although the S/N of the broad H$\beta$ line is significantly
lower than that of the broad H$\alpha$ line, we were able to estimate the broad H$\beta$ flux by setting the width
and center of this line equal to that of H$\alpha$ during fitting.  The Balmer decrements are shown in Table~3 for the broad and narrow
components of Tycho Knot g (1).  For comparison, we have computed $I_{H\alpha}/I_{H\beta}$ for both upper and lower limits of the
visual extinction parameter (1.6\,$\leq\,A_{V}\,\leq$\,2.6) determined by CKR80.
It is evident that the narrow Balmer decrement is considerably larger than the broad Balmer decrement.  Interestingly,
using $A_{V}\,=\,$1.6 yields a broad Balmer decrement of 3.67, consistent with pure collisional excitation in a 
high temperature ($\gtrsim\,$10$^{6}$~K) gas.  Adopting the smaller value of $A_{V}$ also results in better agreement between
observations and models of the photoionization precursor (Ghavamian \etal\, 2000). 

\subsection{RCW 86}

We observed the galactic SNR RCW 86 on April 5-8, 1998 (UT), using the
RC Spectrograph at the f/7.8 focus of the CTIO 4 meter telescope.  The RC Spectrograph features a Loral 3k CCD with
15$\micron$ pixels connected to a Blue Air Schmidt camera.  This combination gives a plate scale
of 0.5$^{\arcsec}$ pixel$^{-1}$.  A decker allows for an adjustable slit length,
with a maximum unvignetted slit size of 5$^{\arcmin}$.  Our observations targeted 4 positions around the
remnant, and included moderate resolution spectra of the H$\alpha$ and
H$\beta$ lines.  In the H$\alpha$ observations, we used the CTIO 1200 l mm$^{-1}$ grating (0.5 \AA\, pixel$^{-1}$, 
blazed at 8000 \AA) with a filter inserted to exclude higher order emission lines.  With a
3$^{\arcsec}$ slit, the spectral resolution of the
H$\alpha$ observations was 2.2 \AA, with spectral coverage of 1565 \AA.   To obtain the H$\beta$ line profiles, we used 
the 860 l mm$^{-1}$ grating in 2nd order (0.34 \AA\, pixel$^{-1}$, blazed at 5500 \AA), with filter bg39 to exclude higher
order emission lines.  Combined with a slit size of 3$^{\arcsec}$, the resolution of the H$\beta$ setup was
1.4 \AA, with spectral coverage of 1135 \AA.

The brightest nonradiative shocks in RCW 86 are located in the southwestern corner of the remnant, flanking the bright
radiative shocks (Figure 5).  From the two-dimensional spectrum of SW RCW 86, it is evident that
some of the shocks at this location have become radiative (marked by the absence of broad Balmer emission and
the presence of [N~II] $\lambda\lambda$6548, 6583 and [S~II] $\lambda\lambda$6716, 6731 lines).
We centered the extraction aperture as shown in Figure 5 to avoid
the radiative emission above and bright stellar continuum below the aperture.  Both optical 
(Leibowitz \& Danziger 1983, Rosado \etal\, 1996, Smith 1997)) and X-ray (Kaastra \etal\, 1992, Smith 1997,
Vink, Kaastra \& Bleeker 1998, Petruk 1999)
observations have shown that the radiative emission arises from a dense cloud overrun by the supernova blast
wave.  The high neutral density is also responsible for the brightness of the broad and narrow H$\alpha$ lines
in the one-dimensional spectrum (Figure 6).

We reduced all spectroscopic data using standard routines in IRAF\footnote{IRAF is distributed by 
the National Optical Astronomy Observatories,
which is operated by the AURA, Inc. under cooperative agreement with the National Science Foundation}.  After 
applying overscan, bias, flat field, response and dark count corrections to all two-dimensional spectra, we
corrected for the slit function
in the two-dimensional spectra using twilight sky flights.  Finally, we
untilted the emission lines using wavelength solutions from calibration lamp spectra, then subtracted the
sky background using night sky emission adjacent to each target object.

The spectra presented in this paper were acquired under varying photometric conditions.
Skies where non-photometric during the 1997 observations, so we only applied a sensitivity
correction to the Cygnus Loop and Tycho data.  Throughout the 1998 observations, however,
conditions at Mt. Hopkins were nearly photometric.  Comparison of the spectra from standard stars Hiltner 102,
LB 1240, HD 192281 and BD 284211 with one another indicates that the 1998 Cygnus Loop spectrum
is photometrically accurate to within 20\%.  The H$\alpha$ observations of RCW 86 were performed under 
nearly photometric conditions,
while the H$\beta$ observations were performed under partial cirrus.  From an examination of spectra from standard stars
LTT 3218, LTT 7379 and LTT 7987, we estimate that the H$\alpha$ flux shown in Figure 6 is accurate 
to within 10\%.

\section{Line Profile Fits}

To extract the broad and narrow component fluxes and measure the width of the broad component, we fit each Balmer line
profile with two independent Gaussians plus a linear background.  Using the IRAF deblending task SPLOT and
self-written routines for $\chi^{2}$ minimization fitting,
we obtained quantitative estimates of the line flux, width and center of the broad and
narrow Balmer lines.  Since the position of the zero level background affects the estimated
flux of the broad component, this baseline uncertainty is the dominant source of
nonrandom error in cases where the broad component is very wide and faint (as in Tycho, for example).  
Both the baseline uncertainty and statistical uncertainty have been included in the quoted broad to narrow ratios.

We measured the broad component widths from the H$\alpha$ line profiles, then corrected for the instrumental response by
subtracting the narrow component width in quadrature from the broad component width.  
In each case, the narrow component widths were equal to the instrumental resolution (they were unresolved).
The broad and narrow components are most tightly blended in the Cygnus P7 line profiles.  
The error bars on the H$\alpha$ and H$\beta$ broad to narrow ratios are correspondingly larger.

\section{The Numerical Shock Models}

The goal of the numerical models is to calculate the H$\alpha$ and H$\beta$ broad to narrow ratios
for arbitrary shock speed, equilibration and preshock neutral density.  Comparing the predicted broad to narrow ratios
with the observed values should then allow us to constrain the shock velocity and equilibration of the observed SNRs.
To compute the Balmer line flux from
a nonradiative shock, we have computed the density and temperature of the postshock gas as a function
of position.  We have used the results of these calculations to compute the Ly $\beta$ and Ly $\gamma$ optical depths
behind the shock. 
 The Lyman optical depth behind the shock typically lies between 0 and 1, meaning that neither Case A
nor Case B conditions apply.  

\subsection{Cross Sections}

We have consulted the following sources for collision cross sections and strengths:

{\it Collisional Ionization and Charge Exchange:} In the shock
models we use the polynomial fit to the electron-hydrogen ionization cross section from Janev \etal\,
(1987).  The proton ionization cross section is from a numerical fit by Janev \etal\, that
reproduces the experimental data of Shah \etal\, (1998), Shah, Elliott \& Gilbody (1987) and
Shah \& Gilbody (1981) to within their uncertainties.   The H$-$H$^{+}$ charge exchange rate used in the
shock code has been taken from the analytic fit of Freeman \& Jones (1974).

{\it Collisional Excitation:}
For electron temperatures below 5$\times$10$^{5}$~K, we use the polynomial fits of Giovanardi, Natta \& Palla
(1987) to the 3$s,p,d$ and 4$s,p,d,f$ collision strengths.  Above 5$\times$10$^{5}$K, we compute the collision rates
directly using the modified Born approximation cross sections of Whelan (1986). The available data
on proton excitation is relatively sparse in the literature, and theoretical cross sections
to $n\,=\,$3 and 4 have only recently become available.  The close-coupling calculations by
Mart\'{\i}n (1999) include cross sections to fine structure levels of $n\,=\,$3 and 4 for proton energies
above 40 keV.  The calculations of Mart\'{\i}n include charge exchange into excited states, a process which
contributes as much as 20\% to the Balmer line flux at high shock velocities ($v_{S}\,\gtrsim\,$1500 \kms).  
The theoretical values agree reasonably well with the experimental measurements of
Detleffsen \etal\, (1994), which covered proton energies in the range 40 keV$\,\leq\,E\,\leq\,$800 keV.
The shock code utilizes the cross sections of Mart\'{\i}n (1999) and uses the calculations of McLaughlin \etal\, (1998)
to estimate excitation cross sections between 10 and 40 keV.

\subsection{Ionization Structures}

The shock structure calculation is initiated by using the jump conditions to set the temperature
and density at the first time step.  
The equations describing the number densities of slow neutrals, fast neutrals,
electrons and protons form a set of coupled, linear differential equations which can be solved together
to calculate the density of each species behind the shock.  Assuming a pure H plasma ($n_{e}\,=\,n_{p}$), 
the equation for the slow neutrals is
\begin{equation}
\frac{d n_{H^{0}}(s)}{d t} \, = \,-\,  n_{e}\,n_{H^{0}}(s)\,(\langle \sigma_{i} v \rangle_{e}^{s} \, + \,
              \langle \sigma_{i} v \rangle_{p}^{s} \, + \, \langle \sigma_{cx} v \rangle^{s})
\end{equation}
where $\langle \sigma_{i} v \rangle_{e,p}^{s}$ and $\langle \sigma_{cx} v \rangle^{s}$ are the electron/proton
ionization coefficients and H$-$H$^{+}$ charge exchange coefficients for slow neutrals (in cm$^{3}$ s$^{-1}$), and
$n$ is the number density (cm$^{-3}$) of a given particle species.
Solving this equation for $n_{H^{0}}(s)$, the slow neutral density at time step $t_{j}\,(=\,t_{j-1}\,+\Delta t)$ is
\begin{equation}
n_{H^{0}}^{j}(s) \, =\, n_{H^{0}}^{j-1}(s)\,e^{- n_{e}^{j-1}\,(\langle \sigma_{i} v \rangle^{s} \, +\,
                \langle \sigma_{cx} v \rangle^{s})\,\Delta\,t}
\label{nhs}
\end{equation}
where $\langle\,\sigma_{i} v\rangle^{s}\,=\,\langle\,\,\rangle_{e}\,+\,\langle\,\,\rangle_{p}$ is the total ionization
coefficient due to protons and electrons evaluated at $t_{j}$.  The relation for fast neutrals is
\begin{equation}
n_{H^{0}}^{j}(f) \, = \, n_{H^{0}}^{j-1}(f)\,e^{- (n_{e}^{j-1}\,\langle \sigma_{i} v \rangle^{f}\,\Delta\,t)} \,
+ \, 4\,n_{H^{0}}^{j-1}(s)\,
e^{- (n_{e}^{j-1}\,\langle \sigma_{i} v \rangle^{s}\,\Delta\,t)}\,(1\, - \, e^{-(n_{e}^{j-1}\,\langle \sigma_{cx} v \rangle
^{s}\,\Delta\,t)})
\label{nhf}
\end{equation}
where $\langle \,\rangle^{f}$ represents the collision rate coefficient for electrons or protons and fast neutrals.
In the above equation, the first term represents the loss of fast neutrals by ionization, while the
second term represents the increase of fast neutrals by charge exchange (the factor of 4 takes into
account the compression behind a strong shock).  Mass conservation requires that the electron and proton
number densities increase in proportion to the ionization rate; therefore the equations for
$n_{H^{0}}(s)$ and $n_{H^{0}}(f)$ can be used to calculate the proton density:
\begin{equation}
n_{p}^{j}\,=\,n_{e}^{j}\,=n_{p}^{j-1}\,+\,4\,n_{H^{0}}^{j-1}(s)\,(1\,-\,e^{- (n_{e}^{j-1}\,\langle \sigma_{i} v \rangle^{s} \Delta\,t)})\, +
\, n_{H^{0}}^{j-1}(f)\,(1\,-\,e^{- (n_{e}^{j-1}\,\langle \sigma_{i} v \rangle^{f} \Delta\,t)})
\label{nep}
\end{equation}

Two important complications arise in the collision rate calculations for a nonradiative shock.
First, $T_{e}\,\neq\,T_{p}$ just behind the shock so that the electron and proton temperatures change
continually with position as Coulomb collisions attempt to establish equilibrium.
Hence, the collision rate coefficients also evolve with position and must be 
computed for a range of electron and proton temperatures. Second, since the slow neutrals
are unaffected by the shock, they encounter an anisotropic proton distribution which is bulk shifted
to $\case{3}{4}\,v_{S}$ (CKR80, Smith \etal\, 1991).  Slow neutrals can also encounter
an anisotropic electron distribution if the plasma is far from temperature equilibrium and the postshock bulk velocity
is comparable to the electron thermal speed.

Another important complication occurs when the temperatures of the electrons and fast neutrals are unequal.  For the
case $T_{e}\,=\,T_{p}$ (equilibrated plasma), the electron thermal speed is 43 times higher than the fast neutral
thermal speed (= proton thermal speed).  The fast neutrals are effectively at rest relative to the electrons,
and the ionization rate integral is simple to evaluate.  However, for an unequilibrated
case, the collision rate integral involves
the distribution functions of both electrons and fast neutrals, and becomes a complicated 6-dimensional
integral over the velocities of both species (Smith \etal\, 1991).  Fortunately, the electron-fast neutral
collision rate integrals can be reduced to one-dimensional form involving the relative speed of the two particle
species (Bandiera, 1998; Weisheit, 1998).
Throughout the shock code, we have computed the electron-fast neutral and electron-slow neutral
rates using the reduced integral.

As mentioned earlier, electrons and protons behind a nonradiative shock can be heated to very different
fractions of the bulk flow energy.  The total thermal energy $E_{tot}$ acquired by the postshock gas is, however,
constant:
\begin{equation}
E_{tot}\,=\,\frac{3}{16}\,(m_{e}\,+\,m_{p})\,v_{S}^{2}\,\approx\,\frac{3}{16}\,m_{p}\,v_{S}^{2}
\end{equation}
The {\it initial} fraction of this energy going into electrons and protons is controlled by collisionless heating
at the shock front, and is therefore a free parameter.  Let us define the parameter $f_{eq}$
such that $f_{eq}\,=\,$0 for no initial electron-proton equilibration and $f_{eq}\,=\,$1 for
full initial electron-proton equilibration.  The postshock temperatures for arbitrary $f_{eq}$ are then
\begin{equation}
T_{p}\,=\,T_{0p}\,+\,\frac{1}{2}\,\frac{3}{16}\,(m_{e}\,f_{eq}\,+\,(2\,-\,f_{eq})\,m_{p})\frac{v_{S}^{2}}{k}
\label{ti}
\end{equation}
and
\begin{equation}
T_{e}\,=\,T_{0e}\,+\,\frac{1}{2}\,\frac{3}{16}\,(m_{p}\,f_{eq}\,+\,(2\,-\,f_{eq})\,m_{e})\frac{v_{S}^{2}}{k}
\label{te}
\end{equation}
where $T_{0p,e}$ are the preshock proton and electron temperatures, respectively. For the extreme cases of no 
equilibration and full equilibration,
\begin{equation}
T_{p,e}\,=\,T_{0p,e}\,+\,\frac{3}{16}\,\frac{m_{p,e}}{k}\,v_{S}^{2}\hspace{0.2in} (f_{eq}\,=\,0)
\end{equation}
and
\begin{equation}
T_{p,e}\,=\,T_{0p,e}\,+\,\frac{1}{2}\,\frac{3}{16}\,\frac{(m_{e}\,+\,m_{p})}{k}\,v_{S}^{2}\,
\approx\,
T_{0p,e}\,+\,\frac{1}{2}\,\frac{3}{16}\,\frac{m_{p}}{k}\,v_{S}^{2}\hspace{0.2in} (f_{eq}\,=\,1)
\end{equation}
The temperature of the preshock medium
is generally low enough so that $T_{0p}\,=\,T_{0e}$\,($=\,T_{0}$).  The initial ratio of $T_{e}/T_{p}$ 
behind the shock can be obtained by dividing Equations \ref{te}\, and \ref{ti}.  Assuming that
$T_{0p}$ and $T_{0e}$ are small enough to ignore, 
\begin{equation}
\frac{T_{e}}{T_{p}}\,\approx\,\frac{f_{eq}}{2\,-\,f_{eq}}
\label{teti}
\end{equation}
where terms involving $m_{e}/m_{p}$ have been dropped.  A diagram showing the dependence of $T_{e}$ and $T_{p}$
on $v_{S}$ and $f_{eq}$ appears in Figure 7.  

To illustrate the dependence of postshock ionization structures on the shock velocity and equilibration, 
we now describe four numerical models (assuming preshock parameters $n_{0}\,=\,$1 cm$^{-3}$, $f_{H^{0}}\,=\,$0.5,
$T_{0}\,=\,$5,000~K):

{\it 250 and 500 \kms\,}: The postshock electron and proton temperatures equilibrate
rapidly in the 250 \kms\, model, even when $f_{eq}\,=\,$0.  The protons are too cold to appreciably ionize H, so charge transfer
and electron collisions dominate the ionization balance of the postshock gas.  The slow and fast neutral densities
are shown as a function of position behind the shock in Figure 8.  Coulomb collisions in the 500 \kms, $f_{eq}\,=\,$0 model
only produce 30\% equilibration by the time the neutrals fully ionize,
and once again charge exchange is the dominant interaction between protons and neutral atoms.  In this model, the broad and narrow
Balmer emission is produced well before equilibration is complete.  Note that the ionization zones for both 250 \kms\, and
500 \kms\, models are more extended for low equilibration than for high equilibration.  At higher equilibrations
the electrons are hotter, ionizing the gas more efficiently and reducing the thickness of the
postshock neutral layer.  At fixed equilibration, the size of the neutral layer is proportional to
the shock velocity. Therefore, the faster the flow, the larger the neutral layer.

{\it 1500 and 2500 \kms\,}: 
There are two important differences between these models and those at lower shock velocity.  First, $T_{p}$ in
a high velocity $f_{eq}\,=\,$0 shock starts off so much higher
than $T_{e}$ that electron-proton equilibration is only a few percent complete by the time all the neutrals disappear
and all the Balmer emission is produced.  Second, the proton ionization rate is now comparable to the electron ionization
rate, and a sizable fraction of the Balmer line emission is produced by proton excitation.
At these shock velocities, the proton-slow neutral collision rate is a factor of 2 or more higher than the
proton-fast neutral collision rate, due to the bulk velocity-shifted proton distribution encountered by the slow neutrals.
At fixed shock velocity, the size of the postshock ionization layer is larger for
high equilibration than low equilibration (Figure~8), the opposite of the 250 and 500 \kms\, models.

\subsection{Calculation of Broad FWHM vs. Shock Velocity}

To make the best use of the observational data, we have used the broad component H$\alpha$ widths in
each SNR to bracket the range of possible shock velocities.  
For a given broad component width, no equilibration
($f_{eq}\,=\,$0) yields the smallest shock velocity $v_{S}(min)$, while full equilibration ($f_{eq}\,=\,$1) 
yields the largest shock velocity $v_{max}$ (CKR80, Smith 1991).  Intermediate 
equilibrations yield intermediate $v_{S}$.  For each observed shock, we ran numerical models sampling a range of 
$f_{eq}$ between 0 and 1, with each value of $f_{eq}$ mapping onto a unique $v_{S}$.  We then compared the
output H$\alpha$ and H$\beta$ broad to narrow ratios with the observed values.  In this manner, our models
self consistently utilized all of the relevant observables from each Balmer line spectrum.

For a given broad FWHM, the map between $f_{eq}$ and $v_{S}$ is the line profile
function $\phi(v_{x})$ (Smith \etal\, 1991, CKR80).  For a plane
parallel shock viewed edgewise, $\phi(v_{x})$ is defined as the number of fast neutrals with velocity $v_{x}$ along
the line of sight:

\begin{equation}
\phi(v_{x})\,=\,\frac{l_{p}^{3}}{\pi^{3/2}\,(\langle \sigma_{cx} v \rangle^{s} + \langle \sigma_{i}
v \rangle^{s})}\,
e^{-l^{2}v_{0}^{2}}\,\int_{-\infty}^{\infty}\,\int_{-\infty}^{\infty}\,v\,\sigma_{cx}(v)\,e^{-l^{2}\,
(v^{2}\,-\,2\,v_{0}\,v_{z})}\,dv_{y}\,dv_{z}
\label{phiv}
\end{equation}

where $v^{2}\,=\,v_{x}^{2}\,+\,v_{y}^{2}\,+\,(v_{z}\,-\,v_{0})^{2}$,
$l_{p}\,=\,(m_{p}/2 kT_{p})^{1/2}$ and $v_{0}\,=\,\case{3}{4}v_{S}$ is the bulk velocity of the
shocked gas, taken to be along the $z$ axis.  We have computed $\phi(v_{x})$ numerically for a range
of $f_{eq}$ and $v_{S}$.   In Figure 9 we present plots of the expected FWHM vs. shock velocity for four
equilibrations.  We have used these results to estimate $v_{S}$ for the observed nonradiative shocks,
listed in Table~4 for the limits of no equilibration and full equilibration.

\subsection{Monte Carlo Models of Ly $\beta$ and Ly $\gamma$ Trapping}

With the densities and temperatures of different particle species known for a given $v_{S}$ and $f_{eq}$, we can
now quantitatively estimate the contribution of Lyman line trapping to the H$\alpha$ and H$\beta$ lines.
Since Lyman photons are emitted with Doppler shifts randomly distributed over the line profile, 
it is possible for a Lyman photon generated near line center of a fast neutral to be absorbed by a slow neutral,
and vice versa.  The conversion of Lyman line photons to Balmer line photons depends on the likelihood
of absorption.  Due to the $\case{3}{4}\,v_{S}$ velocity shift between fast and slow neutrals, the optical
depth depends on both the frequency and direction of an emitted Lyman photon.  Therefore, our
radiative transfer calculation follows the propagation, absorption and conversion of individual Lyman photons
behind the shock.  

We model the conversion of Lyman photons into Balmer photons using a Monte Carlo simulation.  Lyman photons are
propagated through the shock until they are either converted into Balmer photons or escape from the grid.  Due to the
large Doppler motions of fast neutrals, most of the broad Lyman photons escape from the shock.   Narrow
Lyman photons, on the other hand, can be absorbed both behind and ahead of the shock.
There is effectively an infinite optical depth to narrow component Lyman photons in the preshock region; 
some absorption occurs even if the preshock neutral density is very small (CKR80).
For this reason, the radiative transfer calculation includes the preshock
region, where the temperature and density are assumed constant.

The Monte Carlo program used in this work is based on an earlier code used by HRB94, with
atomic data updated to include H$\beta$ emission and proton excitation.  The probability that a Lyman photon will be absorbed
at each point in the grid depends on the optical depth $\tau_{\nu}$ at each point.  The optical depth, in turn, depends on
the random velocity of the emitting atom along the direction of emission, and whether the atom is a fast or slow
neutral.  In the Monte Carlo program, 10,000 individual photons are generated with random Doppler shifts, randomly 
distributed behind the shock according to the emissivity.  If a given excitation produces a Lyman photon, the photon 
is followed along a randomly 
oriented ray until it is either converted into a Balmer photon or escapes from the grid.  After 10,000 excitations, 
the number of accumulated broad and narrow Balmer photons is divided to obtain the broad to narrow intensity ratio.

Before we compare our models with the observations, we note that our predicted broad to narrow ratios 
offer a number of improvements over previous calculations (such as those of HRB94).  The current models utilize more recent
atomic rates and compute the broad to narrow ratios for both H$\alpha$ and H$\beta$.  Our new models also
include direct collisional excitation of hydrogen by protons.  Finally, by arranging our models into a grid we have been
able to predict broad to narrow ratios for a single set of models covering a wide range of shock velocities and
equilibrations.

\section{Comparison of Models and Observations}

With the numerical models available, we can now estimate $v_{S}$ and $f_{eq}$ for
Cygnus P7, SW RCW 86 and Tycho Knot g (1).  
There is observational evidence that the temperature ahead of nonradiative
shocks can be as high as 40,000~K (Ghavamian \etal\, 2000, Smith \etal\, 1994, HRB94).
The model predictions are rather insensitive to the preshock temperature (the Lyman optical depth at line center
$\propto\,T^{-1/2}$), so in the following sections we have set $T_{0}\,=\,$5,000~K in all models. 
It should also be noted that since the preshock neutral density
is equal to the neutral fraction times the total preshock density ($n_{H^{0}}\,=\,f_{H^{0}}\,n_{0}$),
predicted broad to narrow ratios from models with high $f_{H^{0}}$ and low $n_{H^{0}}$ are similar 
to models with low $f_{H^{0}}$ and high $n_{0}$.  The reasons for this property are that (1) the total preshock
density scales out of the ratio of the broad to narrow Lyman line optical depths, and (2) the equilibration,
ionization and charge exchange times behind the shock scale as $1/n_{0}$, offsetting the influence of varying $n_{0}$ on the
excitation/ionization rates.  Along with $v_{S}$ and $f_{eq}$, the preshock neutral fraction $f_{H^{0}}$ is the most important
quantity affecting the broad to narrow ratios.  For this reason, we set $n_{0}\,=\,$ 1 cm$^{-3}$ in the models
which follow.

\subsection{Cygnus Loop P7}

To interpret our results we ran numerical shock models using the fractional equilibrations
$f_{eq}\,=\,$0, 0.03, 0.05, 0.2, 0.5, 0.8 and 1.0, along with their corresponding
shock velocities of 265, 268, 270, 273, 296, 332 and 365 \kms.  
The predicted H$\alpha$ and H$\beta$ broad to narrow ratios are shown in Figure 10
vs. equilibration for preshock ionization fractions of 0.1, 0.5 and 0.9.  In these models, the total preshock
density is 1 cm$^{-3}$.  Each curve in the figure corresponds to a different preshock
neutral fraction $f_{H^{0}}$. Each point along a given curve corresponds to a different equilibration
and maps to a different shock velocity.  An outstanding feature of the models
is that the broad to narrow ratios reach a maximum at low equilibration.  The peak occurs 
because the broad to narrow
ratios are proportional to the ratio of the charge exchange rate to the collisional ionization rate.
In these models the ratio peaks in the range 0.03$\,\lesssim\,f_{eq}\,\lesssim\,$0.05. 
At higher equilibrations the collisional ionization by electrons is increasingly effective, reducing the
size of the neutral layer and reducing the broad to narrow ratio. 

Although variations in $f_{H^{0}}$ do affect the shape and position of the broad to narrow curves,
the most conspicuous feature in Figure 10 is that the broad to narrow ratios are predominantly sensitive 
to $f_{eq}$ (and hence $v_{S}$).  The figure shows that the higher equilibration models
match the observations best.  In H$\alpha$, the predicted broad to narrow ratios fall slightly
above the observed values.  The best match occurs for full equilibration, where the model value lies $\sim\,$10\% above
the upper error bar of the observations.  However, the H$\beta$ broad to narrow ratio does agree with 
observations, favoring $f_{eq}\,\approx\,$0.8$-$1.0 ($T_{e}/T_{p}$ = 0.7$-$1.0 from Equation \ref{teti}).  
Excluding the uncertainty in broad FWHM, the implied range of shock velocities is 300$-$365 \kms.

The disagreement between the observed and predicted H$\alpha$ broad to narrow ratios may be due to the
excitation cross sections utilized in the models.  In the range of shock velocities implied by the observations, the 
thermal energy of the postshock electrons is $\lesssim\,$15 Rydbergs (1 Rydberg = 13.6 eV).  At these energies, the collisional 
excitation cross sections are not as well determined as the cross sections at higher energies. This uncertainty is translated
directly into the calculated broad to narrow ratios. 

The Cygnus P7 shock lies $\sim\,$25$^{\arcmin}$ NW of the filament previously studied by
RBFG83 and HRB94.  In their analysis, HRB94 concluded that their observed shock has recently slowed due
to an encounter with a density enhancement in the ISM.  The shock observed by RBFG83
and HRB94 exhibits an H$\alpha$ broad component width $\sim\,$130 \kms\, and an
H$\alpha$ broad to narrow ratio $\sim\,$1.6.  In addition, faint [O~III], [N~II] and [S~II] emission can
be seen in this filament.  These features are markedly different from those of Cygnus P7.
In a pure Balmer line filament in the western Cygnus Loop, Treffers (1981) also measured an H$\alpha$ broad component 
width $\sim\,$130 \kms, with broad to narrow ratio $\sim\,$1.
Since the broad to narrow ratio is proportional to the ratio of the charge exchange rate to the ionization rate
(CKR80, Smith \etal\, 1991), the difference between the Cygnus P7 broad to narrow ratio and those of other
filaments may be due to the fact that the collisional ionization rate of hydrogen rises strongly with shock 
velocity for 100$\,\lesssim\,v_{S}\,\lesssim\,$300 \kms.  At a given equilibration, this would result in a decreasing broad to 
narrow ratio for this range of shock velocities.

Overall, our derived shock velocity for Cygnus P7 lies well above values obtained in most
other optical studies of the Cygnus Loop.  Previous observations using spectrophotometry
(Raymond \etal\, 1980, RBFG83, Fesen, Kirshner \& Blair 1982, Fesen \& Itoh 1985, Raymond 1988, HRB94),
Fabry Perot spectrometry (Kirshner \& Taylor 1976, Treffers 1981, Shull \& Hippelein 1991) and narrow band imagery 
(Fesen, Kwitter \& Downes
1992, Levenson \etal\, 1998) have detected shock waves in various stages of evolution, from nonradiative to fully radiative.  These
observations have yielded shock velocities $\sim\,$100 \kms\, for the bright, radiative filaments and $\sim\,$150$-$200 \kms\, 
for the faint, partially radiative/nonradiative filaments.   However, Kirshner \& Taylor (1976) and Shull \& Hippelein (1991) 
detected face-on H$\alpha$ emission near the center of the Cygnus Loop, blueshifted to velocities $\sim\,$350$-$400 \kms.  
These authors suggested that the face-on emission is produced by the supernova blast wave propagating into the low density ISM.
Soft X-ray observations of the Cygnus Loop (Ku \etal\, 1984, Levenson \etal\, 1999) confirm the presence of fast ($\sim\,$400
\kms) shocks in this remnant, including locations where there is no optical emission (the preshock gas is fully
ionized).  The Cygnus P7 shock is similar in speed to both the X-ray determined blast wave velocity and the blueshifts seen by Kirshner \&
Taylor (1976) and Shull \& Hippelein (1991).  Therefore, we 
conclude that the Balmer-dominated Cygnus P7 filament marks the current location of the supernova blast wave in the northeastern Cygnus Loop.

\subsection{SW RCW 86}

The H$\alpha$ broad component width of the nonradiative shock RCW 86 SW (Figure 6)
is 562$\pm$18 \kms, placing this shock at a Mach number intermediate between that of Cygnus P7
and Tycho Knot g.  The narrow band H$\alpha$ image (Figure 5) suggests that the shock
geometry here is more complicated than for the Cygnus P7 shock, with multiple filaments surrounding the bright,
clumpy emission covered by the slit.  The forbidden line emission in the two-dimensional spectrum
(Figure 5) indicates that part of the clumpy region has become radiative.  Narrow band
H$\alpha$ and [S~II] images of RCW 86 indicate that most of the bright clumpy emission seen 
in Figure~5 is produced by radiative shock waves (Smith 1997).  From the one-dimensional spectrum 
in Figure~6 we estimate an H$\alpha$ surface brightness of 4.2$\times$10$^{-5}$ \sb\, for the nonradiative shocks
in SW RCW 86.  This surface brightness is nearly twice that of Knot g (KWC87, 
Ghavamian 2000) and a factor of two or more brighter than the filament observed by HRB94. 
This property, along with the presence of radiative
emission nearby, suggests an enhanced preshock neutral density in this part of the SNR.  The velocity shift
between the broad and narrow component centers in Figure 6 is smaller than the uncertainty in broad component width,
indicating that the shock is viewed nearly edge-on.

Broad to narrow ratios were previously measured in RCW 86 by Long \& Blair (1990), who obtained optical spectra of
Balmer-dominated shocks in the N and SW (near our slit position).  In SW RCW 86, they obtained an H$\alpha$ broad 
component width of 600$-$780 \kms\, and a broad
to narrow ratio of 0.65$-$0.94 for that line.  These parameters are noticeably different from ours; the
discrepancy is likely due to the fact that Long \& Blair (1990) observed a partially radiative shock close to
our slit position.  This is evidenced
by the faint [N~II] and [S~II] emission lines in their SW RCW 86 spectrum.  
In this case, slight contamination of the H$\alpha$ narrow component
by radiative H$\alpha$ has likely lowered the broad to narrow ratio.  In addition,
the S/N in the H$\alpha$ profile of RCW 86 SW is much lower in the Long \& Blair (1990) data than ours, making
their broad component line fit more uncertain.  

We assembled a grid of shock models for equilibrations $f_{eq}\,=\,$0, 0.03, 0.05, 0.2, 0.5, 0.8 and
1.0.  These equilibrations yield the observed FWHM for shock velocities of 565, 570,
573, 575, 634, 710 and 775 \kms, respectively.  The predicted broad to narrow ratios
appear in Figure 11, computed for the case $n_{0}\,=\,$1 cm$^{-3}$. 
As with Cygnus P7, models with higher equilibration
match the observations somewhat better than the low equilibration models.
From Figure 11, $f_{eq}\,\approx\,$0.4$-$0.5 ($T_{e}/T_{p}\,\approx\,$0.25$-$0.33) simultaneously
matches the H$\alpha$ and H$\beta$ broad to narrow ratios.  The implied shock velocity is 600$-$635 \kms.

The ability to simultaneously determine both the velocity and equilibration of Balmer-dominated shocks is
a potentially useful aid in modeling the X-ray emission from SNRs.  The ASCA study of RCW 86 by
Borkowski \etal\, (2000) is one particular example.  The Balmer-dominated shocks
in SW RCW 86 lie close to one of the ASCA spectral extraction windows used by Borkowski \etal\, (2000).
In this section of the remnant, the X-ray emission is dominated by nonthermal continuum from fast
($v_{S}\,\sim\,$600$-$1000 \kms) shocks, with thermal emission lines barely detected.  Borkowski \etal\, (2000) 
noted that the portion of SW RCW 86 lying in the ASCA window features a variety of shock speeds and that 
characterizing the thermal and nonthermal emission requires a large number of free parameters.  This makes
the X-ray modeling of SW RCW 86 quite difficult.  We note that in the future, velocities and equilibrations
predicted by our Balmer-dominated shock models may be used to reduce the number of free parameters needed
to model the X-ray data of RCW 86 (and other SNRs).

\subsection{Tycho Knot g}

We compared data from Knot g (1) with shock model predictions for equilibrations of
0, 0.03, 0.05, 0.2, 0.5, 0.8 and 1.0, corresponding to shock velocities of 2050, 2075, 2090,
2175, 2375, 2635 and 2900 \kms.  We allowed the preshock neutral fraction to vary
from 0.99 to 0.1.  We present the predicted broad to narrow ratios for Knot g (1) in Figure 12.
The models are unable to match the observed H$\alpha$ broad to narrow ratio, although the
H$\beta$ broad to narrow ratio is consistent with $f_{eq}\,\lesssim\,$0.2
($T_{e}/T_{p}\,\lesssim\,$0.1).  This is similar to the limit on $T_{e}/T_{p}$ derived by Laming \etal\, (1996)
for a shock of similar speed in SN 1006.  Ignoring the measurement uncertainty in broad component width, the
corresponding shock velocity of Knot g (1) is 2050$-$2175 \kms.
The minimum in $I_{b}/I_{n}$ occurs for an equilibration of 3\%$-$5\%; this corresponds to the
optimum electron temperature for Balmer line excitation in the narrow component
(the opposite of what occurs in the slower shock models).  Models with high preshock neutral fractions
($\gtrsim\,$0.8) match the H$\beta$ broad to narrow ratio best; this result agrees with predictions
of our photoionization precursor models (Ghavamian \etal\, 2000).
The high neutral fraction may be explained by the H~I 21 cm observations of Reynoso \etal\, (1999), 
which suggest that the eastern edge of Tycho is encountering a warm H~I cloud.  On the 
other hand, X-ray observations (Hwang, Hughes \& Petre 1998, Seward 1983) and evolutionary models (Dwarkadas \&
Chevalier 1997) of Tycho's SNR indicate a total preshock density $\lesssim\,$1.1 cm$^{-3}$.  Taken together,
the optical, X-ray and radio data suggest that Knot g is probably encountering the extreme (low density) edge of the H~I
cloud.

\section{Discussion}

One obvious problem is the failure of the models
to reproduce the observed H$\alpha$ broad to narrow ratio for Tycho Knot g (1).  The predicted H$\beta$ broad to narrow ratios
do agree with the observations, although this is due in part to the larger error bars.  
Taking $A_{V}\,=\,$1.6, the predicted narrow component Balmer
decrement falls below the observed value, regardless of the equilibration or Lyman line trapping efficiency.
The disagreement between the observed and predicted broad to narrow ratios may be partly due to the
proton excitation cross sections utilized by the shock code.  The proton
cross sections are most uncertain at low energy,$\sim\,$10 keV.  Since the proton thermal energy in the Knot 
g (1) shock lies close to this value, the H$\alpha$ and H$\beta$ excitation rates 
computed from these cross sections (including charge transfer into excited states)
will be correspondingly uncertain.  This situation is similar to the problem mentioned earlier with the Cygnus P7 models,
where the uncertain quantity was the electron excitation cross section near threshold.

An alternate explanation of the discrepant H$\alpha$ broad to narrow ratio is
collisional excitation of H in a spatially unresolved precursor.  If the upstream H is collisionally excited before
entering the Knot g shock, the resulting H$\alpha$ and H$\beta$ emission
will add to the overall narrow component flux from the shock.  In an earlier paper (Ghavamian \etal\, 2000) we
presented a high resolution H$\alpha$ spectrum which indicated that the neutrals are heated
to a temperature $\sim\,$40,000~K just before entering the shock.  At temperatures $\sim\,$10$^{4}$~K, collisional
excitation produces an optical spectrum with a steep Balmer decrement; this could explain why
the narrow component H$\alpha$ flux is so enhanced relative to that of  H$\beta$.  The Balmer-dominated
spectrum of Tycho would then result from the superposition of precursor + postshock neutral excitation.

It would be appropriate here to speculate on the relationship between $f_{eq}$ and the magnetosonic
Mach number $M_{S}$ ($=\,$$v_{S}/(c_{S}^{2} + v_{A}^{2})^{1/2}$, where $c_{S}$ is the sound speed and
$v_{S}$ is the Alfv\'{e}n speed in the preshock gas).  There is a clear trend of decreasing equilibration with higher
$M_{S}$ (see Table~5).  Cargill \& Papadapoulos (1988) 
have argued that for low Mach number shocks propagating perpendicular to the interstellar magnetic field, a series
of plasma instabilities are initiated by the protons just behind the shock.
As the protons gyrate about the field lines, some of them re-enter the
upstream region, counterstreaming into the preshock gas.  Laming (1998, 2001) and Bingham \etal\, (1997) have pointed out that if
the reflected protons follow a beamlike (monoenergetic) distribution, a two-stream instability
will develop.  Lower hybrid waves generated by the instability can be in resonance with protons and electrons simultaneously,
allowing energy transfer between the two particle species.  A key difference between perpendicular shocks with 
$M_{S}$ $\sim\,$20$-$50 (like Cygnus P7 and SW RCW 86) and those with $M_{S}\,\sim\,$200 like those in Tycho) is that the 
former are expected to be laminar 
(i.e., steady) while the latter are expected to be highly turbulent (Tidman \& Krall 1971).  Due to the unsteady
nature of high Mach number shocks, reflected protons are likely to exhibit an angular spread in velocities,
inhibiting the growth of the two-stream instability and reducing the efficiency of electron heating.  Electron
heating by lower hybrid waves has been used to explain X-ray emission from comet C/Hyakutake (Bingham \etal\,
1997), the nonthermal X-ray tail observed in Cas A spectra (Laming 2001) and the injection of thermal particles
into the cosmic ray acceleration process (McClements \etal\, 1997).

There is strong evidence of an inverse relation between $f_{eq}$ and $M_{S}$
from spacecraft observations of solar wind shocks.  Schwartz \etal\, (1988) presented ISEE data from 14 interplanetary
shocks, as well as 66 crossings of the Earth's bow shock.  They found that across the shock transition electrons 
are heated relative to the protons by an amount which scales as 1/$M_{S}$.  Interestingly, the fractional equilibrations
we derive from our analysis follow roughly the same trend.  The shocks studied by
Schwartz \etal\, only reach $M_{S}\,\sim\,$20, while our observations cover 25$\,\lesssim\,M_{S}\,\lesssim\,$200.  
The equilibration fraction determined by Laming \etal\, (1996) for a nonradiative shock in SN 1006 agrees with the value found 
simply by extrapolating $f_{eq}\,\propto\,1/M_{S}$ from the lower Mach number shocks of Schwartz \etal\, (1988) to the 
values found in SN 1006 ($M_{S}\,\sim\,$250).

It is interesting to note that according to the results of Cargill \& Papadapoulos (1988), collisionless
heating proceeds more efficiently for shocks propagating perpendicular to the interstellar magnetic field.
There may be a connection between the more efficient equilibration predicted for perpendicular shocks and the limb brightening
seen in radio observations of barrel shaped remnants.  In these cases, the limb brightening occurs in parts of the remnant which
propagate perpendicular to the magnetic field (Gaensler 1998).  The radio (and sometimes X-ray) emission from these remnants 
is synchrotron radiation from electrons energized by first order Fermi acceleration, a process where electrons
are boosted to relativistic energies by scattering back and forth between upstream and downstream
turbulence (Jones \& Ellison 1991, Reynolds \& Gilmore 1986, Blandford \& Eichler 1987, Jones \& Ellison 1991).
Fulbright \& Reynolds (1990) proposed that the enhanced radio emission occurs because
the Fermi acceleration process is more efficient for perpendicular shocks.  From the work of McClements \etal\, (1997) and
Laming (1998), it appears that the same plasma instabilities which heat the electrons in strong shocks
can also inject particles into the cosmic ray acceleration process.  Clearly, the 
relationship between magnetic field orientation, equilibration and cosmic ray acceleration 
deserves a more detailed investigation.

\section{Conclusions}

To investigate the properties of nonradiative shocks, we have obtained high S/N spectra of the Cygnus Loop,
RCW 86 and Tycho covering a factor of 10 in Mach number.  In each remnant, we have measured broad to narrow ratios
of both H$\alpha$ and H$\beta$.   We find that the broad to narrow ratios show considerable variation
from one remnant to the next, with the H$\beta$ ratio systematically larger than that of H$\alpha$.  The difference
between the two ratios is evidence of Lyman line trapping in the narrow component.

We have devised a numerical code which predicts the ionization structure behind a nonradiative shock and uses a
Monte Carlo simulation to calculate the influence of Ly $\beta$ and Ly $\gamma$ trapping on the H$\alpha$ and H$\beta$
emission lines.  We have modeled the H$\alpha$ and H$\beta$ broad to
narrow ratios in three of the observed shocks: Cygnus P7 ($v_{S}\,=\,$300$-$400 \kms), SW RCW 86 
($v_{S}\,=\,$580$-$660 \kms) and Tycho Knot g ($v_{S}\,=\,$1940$-$2300 \kms).  Overall, the shock code matches
the observations in the first two remnants, but yields only marginal agreement for Tycho Knot g.  The models
indicate nearly complete equilibration for Cygnus P7 and half equilibration for RCW 86 SW, evidence for substantial
collisionless heating in these nonradiative shocks.  In Knot g, the predicted H$\alpha$ broad to narrow
ratios are systematically larger than the observed values for all equilibrations.
The difficulty in modeling the Knot g broad to narrow ratio may be due to the large uncertainty in proton 
excitation cross sections near threshold.   However, the models do successfully reproduce the observed
H$\beta$ broad to narrow ratios.  
From the H$\beta$ results, we infer a low equilibration for Tycho, $\lesssim\,$20\%.  This value is consistent
with the findings of Laming \etal\, (1996), who used ultraviolet observations with HUT to determine the equilibration of
shocks of similar strength in SN 1006.

For a given shock velocity, we find that the thickness of the ionization layer depends on the electron-proton
temperature equilibration.  At shock velocities $\lesssim\,$1000 \kms, the collisional ionization of
hydrogen is dominated by the electrons; the greater the equilibration, the thinner the ionization
layer.  This trend is reversed at high shock velocities.  Above 2500 \kms, proton ionization dominates electron
ionization; therefore the more equilibrated the shock, the thicker the ionization layer.

P. G. would like to the thank J. Weisheit, J. M. Laming and R. Bandiera for several helpful discussions.
We would also like to thank the anonymous referee for valuable suggestions in improving the presentation of this paper.
The work of P. G. was supported by grant D70832 from Rice University, STScI grant GO-07515-02.96A,
NSF atomic physics theory grant PHY-9772634 and student travel support from CTIO.  P.G. also acknowledges the
hospitality of the Harvard-Smithsonian Center for Astrophysics, where some of this paper was completed.  The work 
of J. R. was supported by NASA grant NAG 5-2845.

\clearpage

\figcaption{Top: Red image from the Palomar Optical Sky Survey, showing the P7 nonradiative shock in 
NE Cygnus Loop.  The approximate position and width of the FAST spectrograph slit is marked.
Bottom: Two-dimensional sky subtracted spectrum of position P7, showing the H$\alpha$ broad and narrow
components.   Extraction aperture for the one-dimensional spectrum is indicated
by a bracket.  East is at the top.}

\figcaption{The extracted one-dimensional H$\alpha$ and H$\beta$ profiles of Cygnus Loop P7.  }

\figcaption{
Top: Narrow band H$\alpha$ image of Tycho, acquired with the direct imager at the 4m telescope at KPNO.
The FAST slit position is indicated.  Bottom: The two-dimensional sky subtracted spectrum of Knot g,
showing broad and narrow H$\alpha$ emission.  The broad component is lost in the noise near the
bottom of the slit, where the emission is faintest.  Extraction aperture for the one-dimensional
spectrum (position 1 in the text) is indicated by a bracket.  North is at the top.
}

\figcaption{The one-dimensional H$\alpha$ and H$\beta$ profiles of Knot g (1).  The spectrum has
not been corrected for interstellar reddening.}

\figcaption{Top: the southwestern portion of RCW 86, from Smith (1997).  The Balmer-dominated filaments covered by
the slit are among the brightest nonradiative shocks observed to date.  Radiative emission can
be seen in the upper right side of the image.  Bottom: the two-dimensional sky subtracted
spectrum of the southwestern filament. The extraction aperture of the one-dimensional spectrum
is indicated by a bracket.  Northwest is located at the top.}

\figcaption{
H$\alpha$ and H$\beta$ line profiles for southwest RCW 86.  No reddening correction has been applied.}

\figcaption{The electron and proton temperatures predicted by the jump conditions, shown for a range of
shock velocities and equilibrations. The calculation assumes a pure H gas. }
          
\figcaption{Ionization structures of four nonradiative shocks, shown for a 50\% preionized medium
with total number density of 1 cm$^{-3}$.  Densities of fast and slow neutrals are plotted vs. postshock distance
for the cases of no equilibration ($f_{eq}\,=\,$0) and full
equilibration ($f_{eq}\,=\,$1) at the shock front.  }

\figcaption{The variation of broad component FWHM with shock velocity, computed for different equilibrations
using Equation \ref{phiv}.  The numerical values in the figure
have been computed for a nonradiative shock viewed edge-on.}

\figcaption{The predicted broad to narrow ratios for Cygnus Loop P7 (broad FWHM = 262$\pm$32 \kms),
shown vs. equilibration fraction $f_{eq}$ for three different preshock neutral fractions ($n_{0}\,=\,$1 cm$^{-3}$). 
Each equilibration combines with a unique shock velocity to yield the observed broad FWHM.
Horizontal dashed lines mark the range of broad to narrow ratios derived from observations.}

\figcaption{The predicted broad to narrow ratios for RCW 86 SW (2) (broad FWHM = 562$\pm$18 \kms),
shown vs. equilibration fraction $f_{eq}$ for three different preshock neutral fractions ($n_{0}\,=\,$1 cm$^{-3}$). 
Each equilibration combines with a unique shock velocity to yield the observed broad FWHM.
Horizontal dashed lines mark the range of broad to narrow ratios derived from observations.}

\figcaption{The predicted broad to narrow ratios for Tycho Knot g (1) (broad FWHM = 1765$\pm$118 \kms),
shown vs. equilibration
fraction $f_{eq}$ for three different preshock neutral fractions ($n_{0}\,=\,$1 cm$^{-3}$).  Each equilibration 
combines with a unique shock velocity to yield the observed broad FWHM.
Horizontal dashed lines mark the range of broad to narrow ratios derived from observations.}

\begin{deluxetable}{ccccccccccccl}
\tablecaption{Journal of Spectroscopic Observations}
\tablehead{
\colhead{Target} &
\colhead{$\alpha$(2000)} & {$\delta$(2000)} & \colhead{Date (UT)} & 
\colhead{Range (\AA)} & \colhead{P.A.\tablenotemark{a}} & \colhead{Exposure (s) } }
\startdata
Cygnus P7 & 20:54:32.5 & 32:17:32.8 & 1997 Oct 7  & 6050-7050 & 90$^{\circ}$ & 6$\times$900\\
Cygnus P7 & 20:54:32.5 & 32:17:32.8 & 1998 Sep 26 & 4370-5370 & 90$^{\circ}$ & 11$\times$1800\\
RCW 86 SW & 14:41:08.1 & $-$62:43:54.5 & 1998 Apr 6 & 5800-7430 & 128.7$^{\circ}$ & 4$\times$900\\
RCW 86 SW & 14:41:08.1 & $-$62:43:54.5 & 1998 Apr 8 & 4610-5110 & 128.7$^{\circ}$ & 3$\times$1800\\
Tycho Knot g & 00:25:50.9  &  64:09:19.2 & 1997 Oct 10 & 3600-7600 & $-$10$^{\circ}$ & 6$\times$1800\\
\enddata  
\tablenotetext{a}{Measured E of N}
\end{deluxetable}

\begin{deluxetable}{ccccccc}
\tablecaption{Present and Prior H$\alpha$ Measurements of the Knot g Broad Component}
\tablehead{
\colhead{Paper} & \colhead{FWHM (\kms)} & 
\colhead{$\Delta\,V_{bn}$ (\kms)\tablenotemark{a}} & \colhead{I$_{b}$/I$_{n}$}  }
\startdata
Chevalier, Kirshner \& Raymond (1980) & 1800$\pm$200 & \nodata & 0.4$-$1.3\\
Kirshner, Winkler \& Chevalier (1987) & 1800$\pm$100 &  238$\pm$18 & 1.08$\pm$0.16 \\
Smith \etal\, (1991) & 1900$\pm$300 &  240$\pm$60 & 0.77$\pm$0.09 \\
Present Work: Knot g (1) & 1765$\pm$110  & 132$\pm$35 & 0.67$\pm$0.1\tablenotemark{b} \\
Present Work: Knot g (2) & 2105$\pm$130 & $<$ 130 & 0.75$\pm$0.1\tablenotemark{b} \\
\enddata  
\tablenotetext{a}{Velocity shift between broad component line center and narrow component}
\tablenotetext{b}{Broad to narrow ratio corrected for contribution from diffuse emission}
\end{deluxetable}

\begin{deluxetable}{ccccccc}
\tablecaption{Tycho Knot g (1) Balmer Decrements}
\tablehead{
\colhead{A$_{V}$\tablenotemark{a}} &
\colhead{H$\alpha$/H$\beta$ (broad)} & \colhead{H$\alpha$/H$\beta$ (narrow)} &
\colhead{H$\alpha$/H$\beta$ (total)} }
\startdata
Undereddened & 6.11$^{\,+8.0}_{\,-2.51}$ & 9.93$^{\,+1.57}_{\,-1.23}$ & 7.97$^{\,+4.47}_{\,-2.27}$\\
& & & & \\
1.6  &  3.67$^{\,+4.83}_{\,-1.5}$ & 5.96$^{\,+0.93}_{\,-0.73}$ & 4.78$^{\,+2.7}_{\,-1.36}$\\
& & & & \\
2.6  &  2.66$^{\,+3.52}_{\,-1.08}$ & 4.33$^{\,+0.68}_{\,-0.53}$ & 3.48$^{\,+1.95}_{\,-0.99}$
\enddata  
\tablenotetext{a}{$A_{V}\,=\,$1.6 and 2.6 correspond to the range of visual extinction
parameters inferred for Tycho (Chevalier, Kirshner \& Raymond 1980).}
\end{deluxetable}

\begin{deluxetable}{ccccccccc}
\tablecaption{Line Profile Fits}
\tablehead{
\colhead{Target} &
\colhead{I$_{b}$/I$_{n}$\,(H$\alpha$)} & \colhead{I$_{b}$/I$_{n}$\,(H$\beta$)} 
& \colhead{Broad FWHM (km/s)} & \colhead{v$_{S}$ (km/s)\tablenotemark{a}} }
\startdata
Cygnus P7 & 0.59$\pm$0.3 & 0.99$\pm$0.3  & 262$\pm$32 & 235$-$395\\
RCW 86 SW & 1.18$\pm$0.03 & 1.54$\pm$0.17 & 562$\pm$18 & 545$-$793\\
Tycho Knot g & 0.67$\pm$0.1\tablenotemark{b} & 1.15$\pm$0.3 & 1765$\pm$110 & 1940$-$3010\\ 

\enddata  
\tablenotetext{a}{Extrema in shock velocity $v_{S}$ correspond to the equilibrated and unequilibrated cases,
respectively. Quoted $v_{S}$ includes uncertainty in broad FWHM.}
\tablenotetext{b}{Corrected for contribution from diffuse emission.} 
\end{deluxetable}

\begin{deluxetable}{ccccccc}
\tablecaption{Shock Parameters Predicted by Numerical Models}
\tablehead{
\colhead{Shock} &
\colhead{$f_{eq}$} & \colhead{$T_{e}/T_{p}$} &
\colhead{$v_{S}$ (km s$^{-1}$)\tablenotemark{a}} }
\startdata
Cygnus P7 & 0.8$-$1.0 & 0.67$-$1.0 & 300$-$400\\
SW RCW 86 & 0.4$-$0.5 & 0.25$-$0.33 & 580$-$660\\
Tycho Knot g (1) & $\leq\,$0.2 & $\leq\,$0.1 & 1940$-$2300\\
\enddata  
\tablenotetext{a}{Includes measurement uncertainty of broad component width.}
\end{deluxetable}

\clearpage

\begin{figure}
\epsscale{0.7}
\plotone{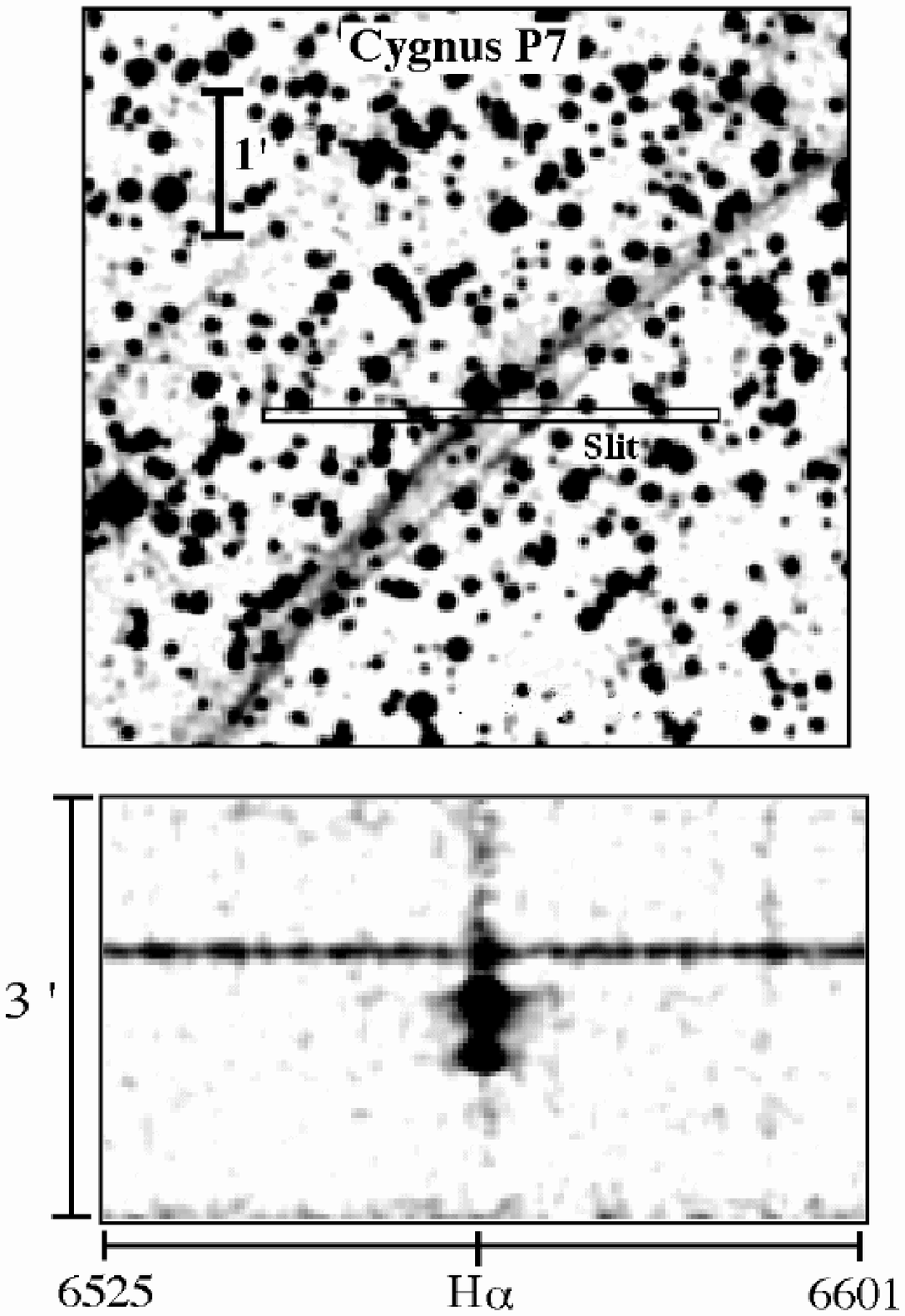}
\epsscale{1.}
\end{figure}

\begin{figure}
\plotone{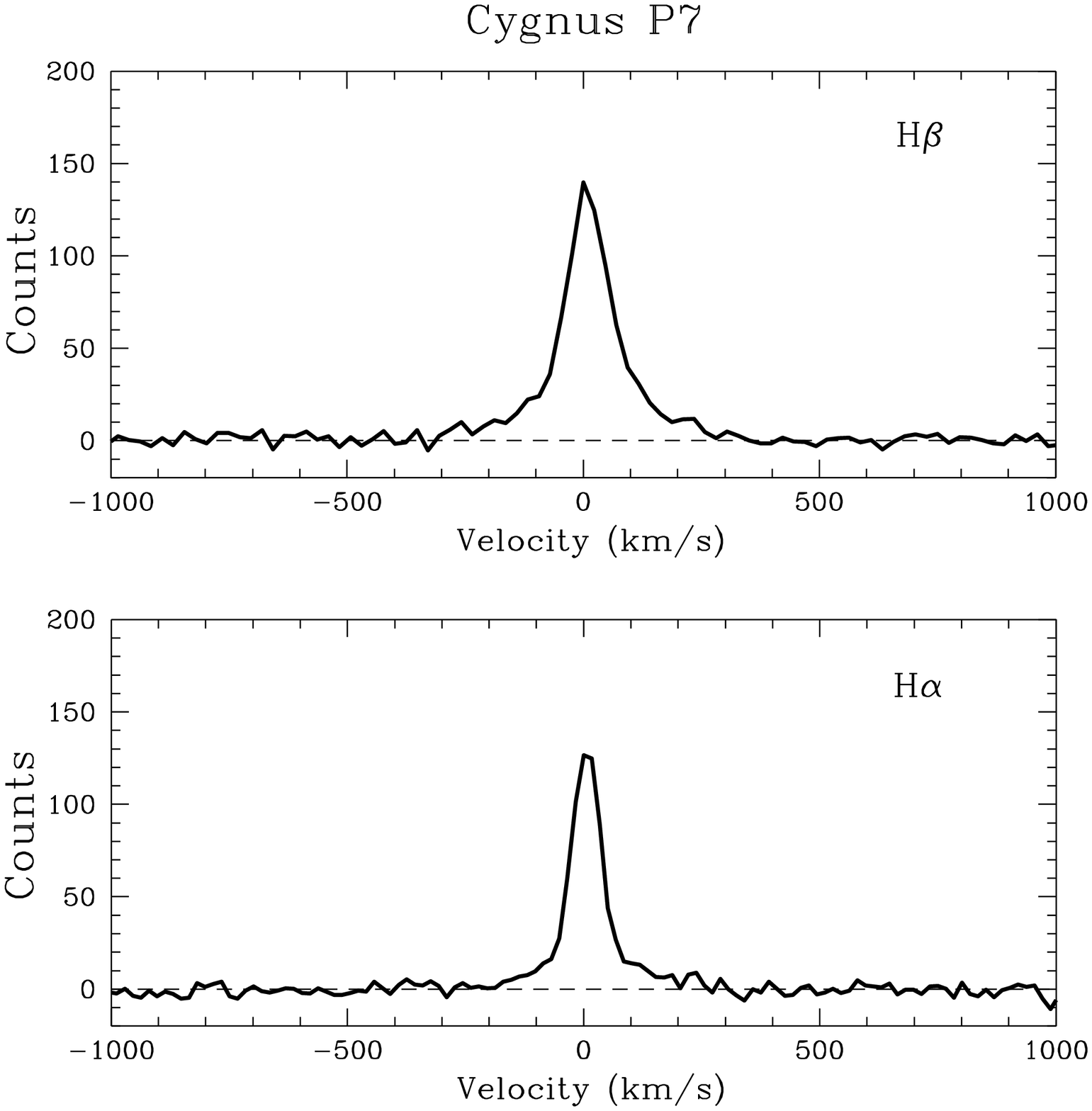}
\end{figure}

\begin{figure}
\plotone{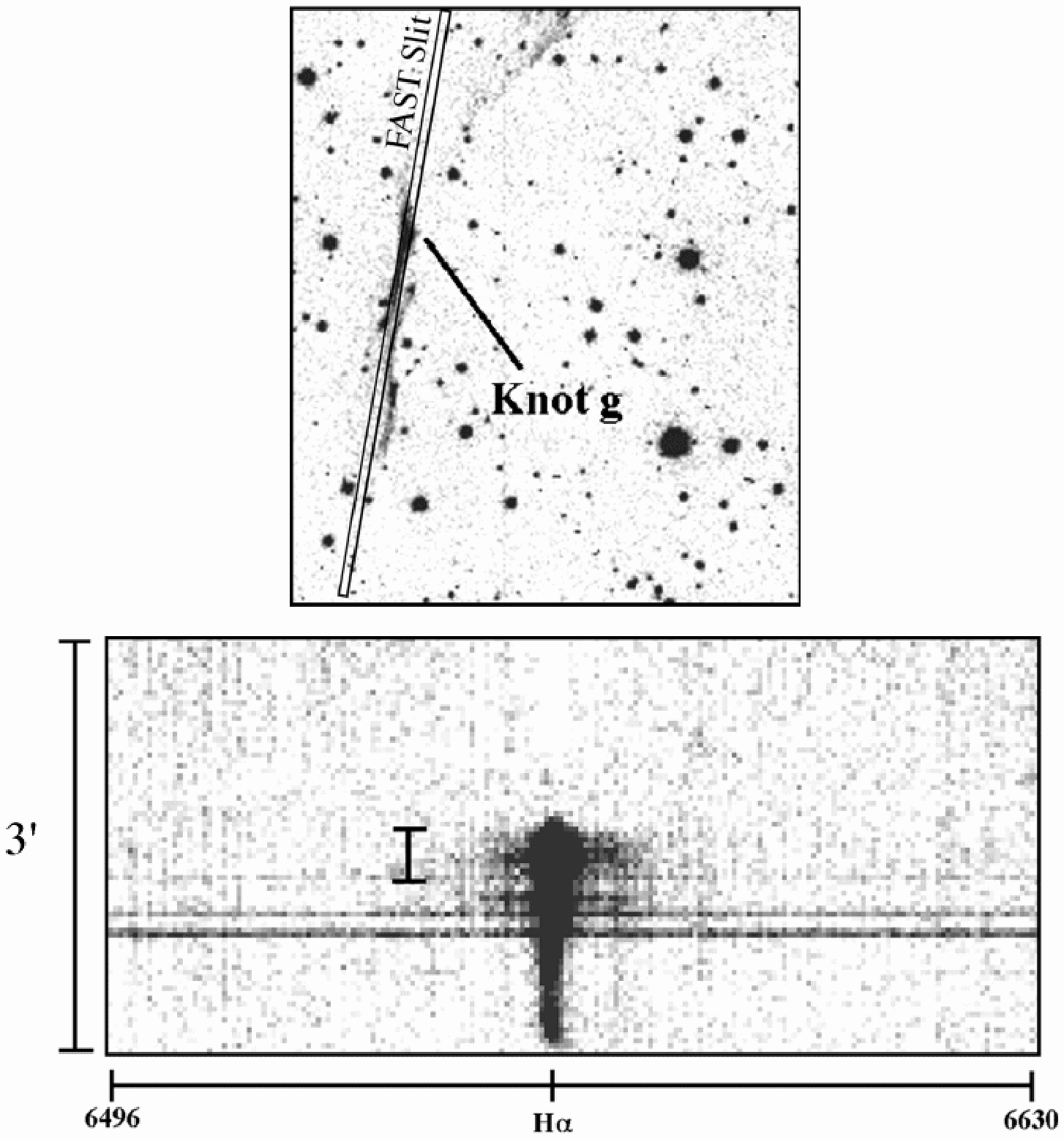}
\end{figure}

\begin{figure}
\plotone{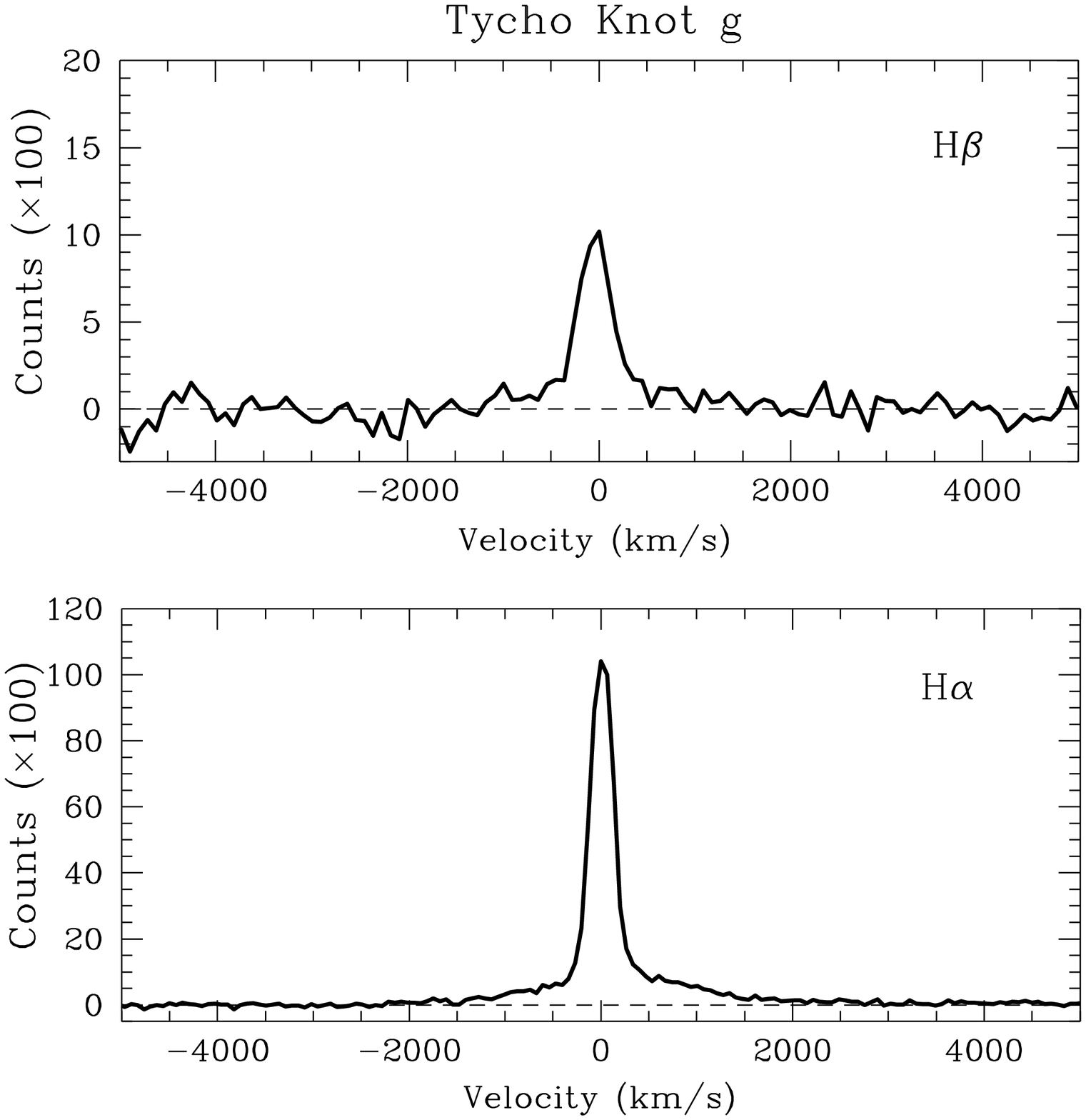}
\end{figure}

\begin{figure}
\epsscale{0.7}
\plotone{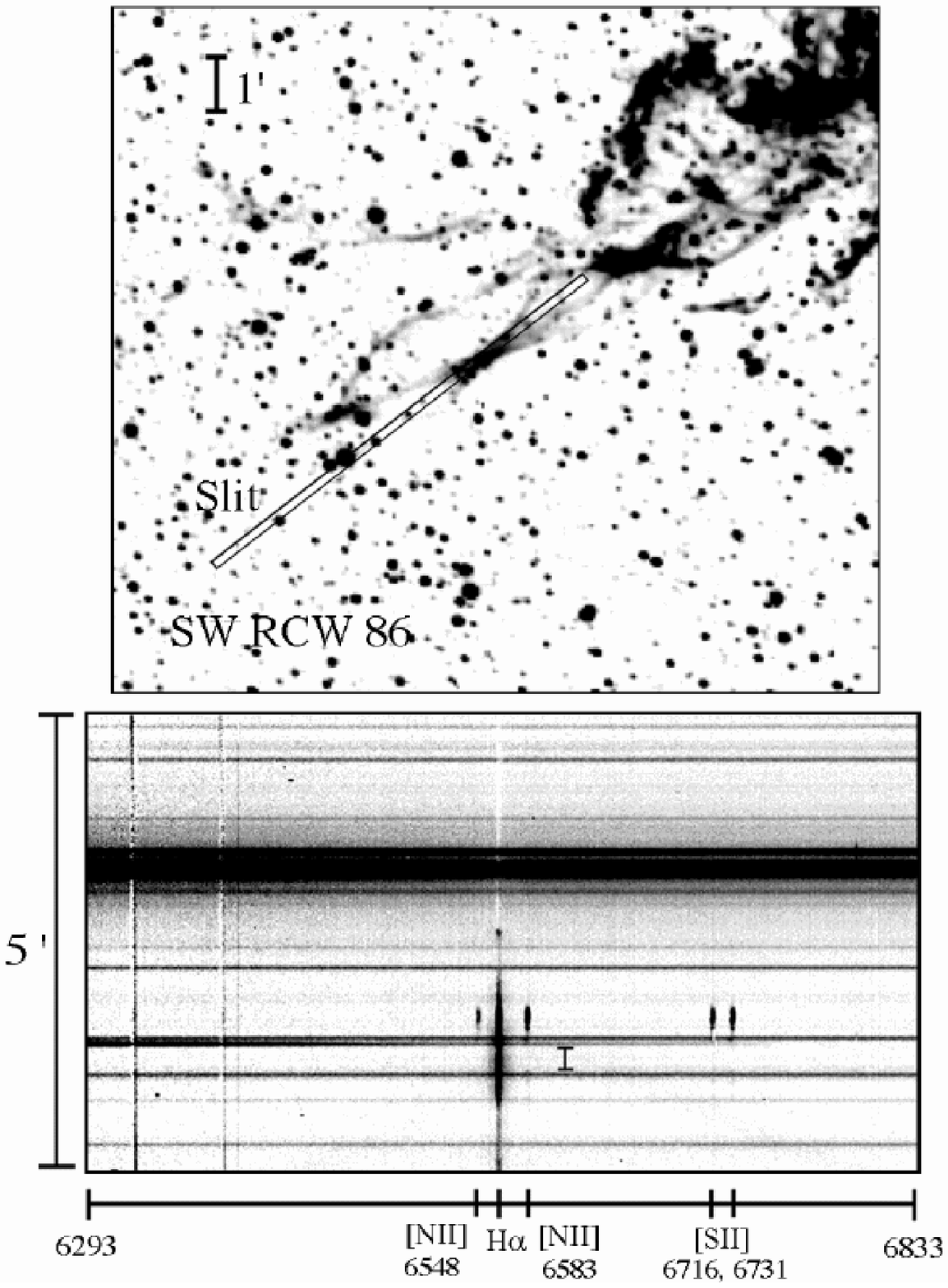}
\epsscale{1.}
\end{figure}

\begin{figure}
\plotone{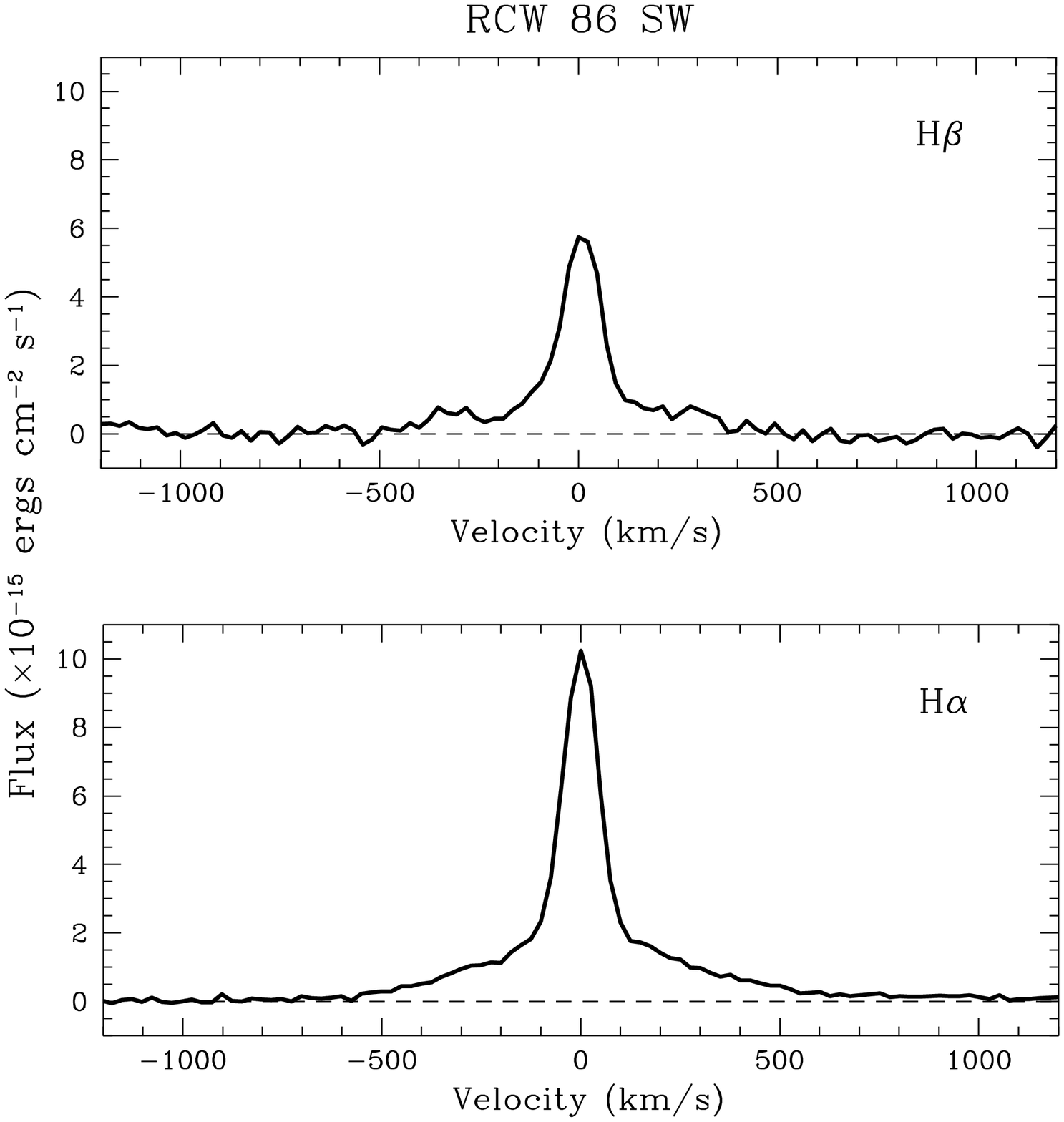}
\end{figure}

\begin{figure}
\plotone{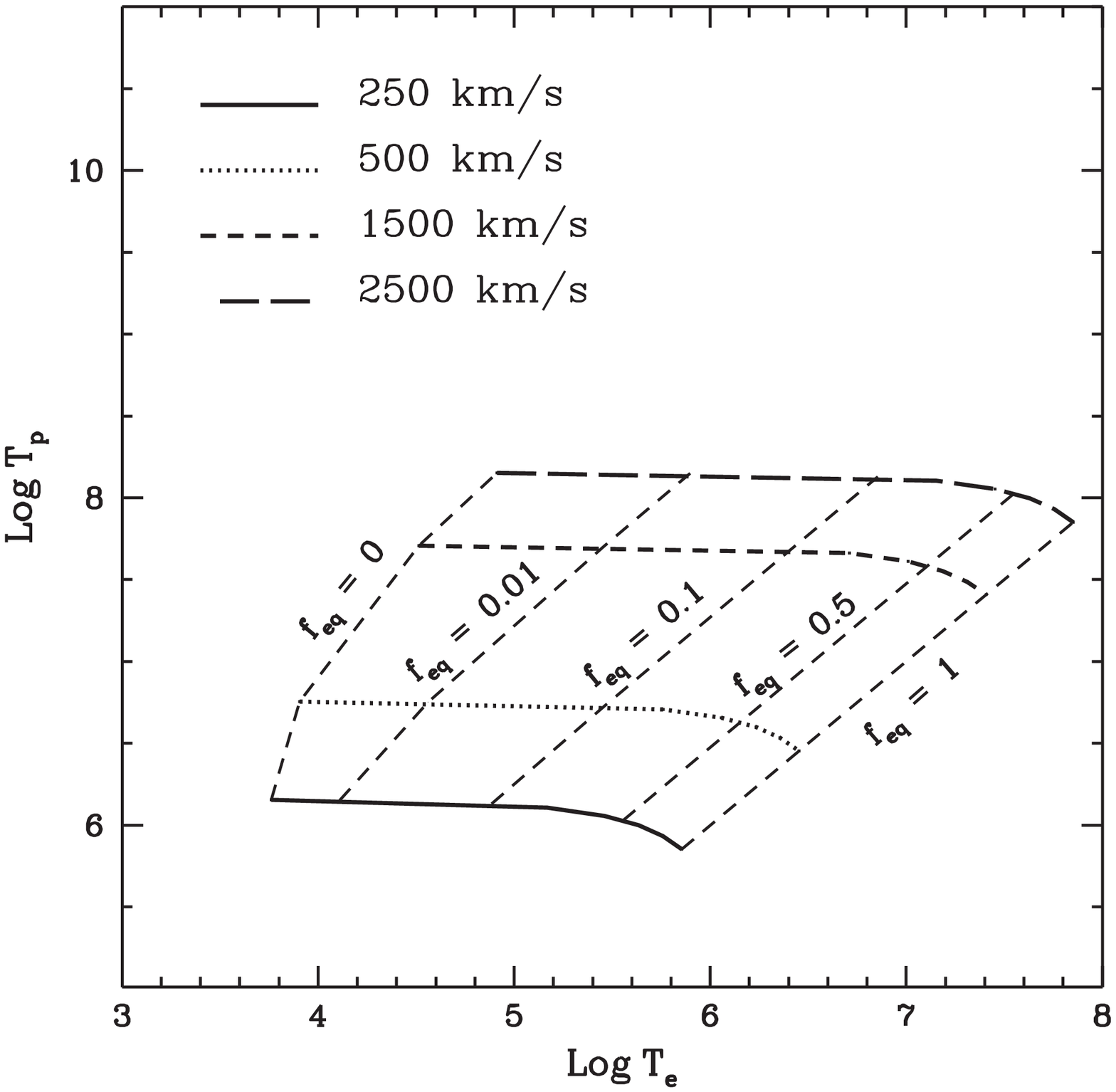}
\end{figure}

\begin{figure}
\plotone{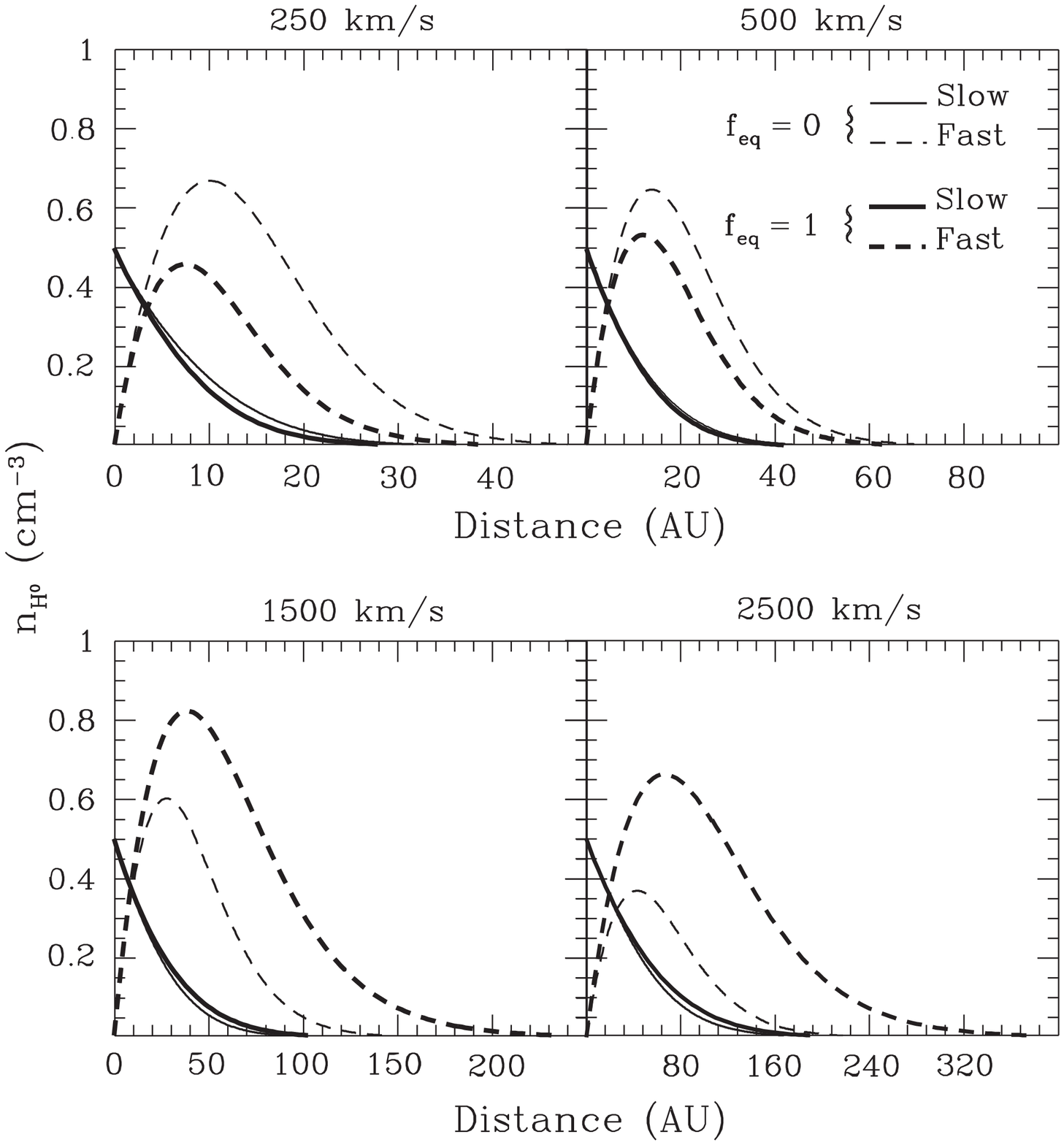}
\end{figure}

\begin{figure}
\plotone{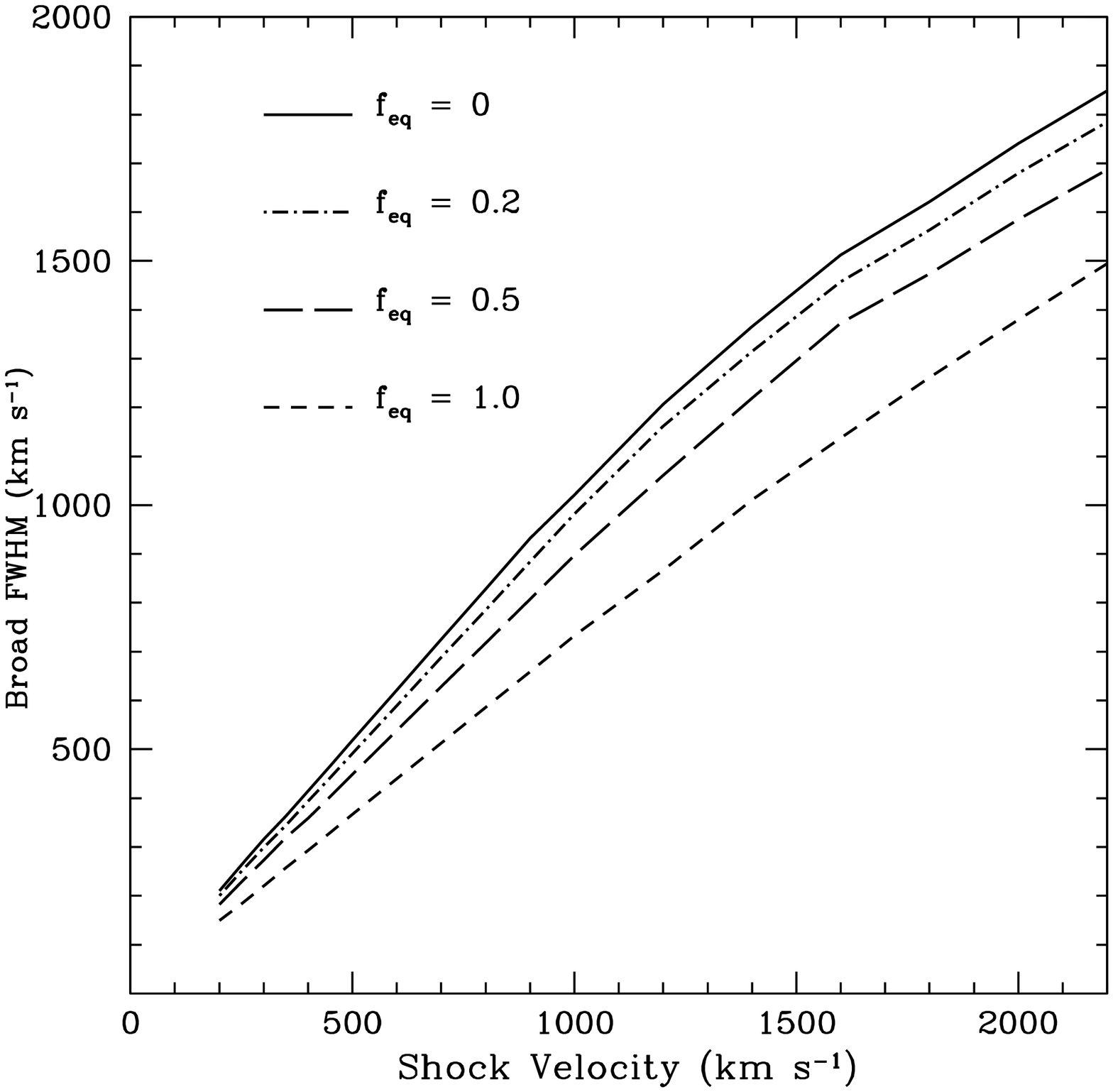}
\end{figure}

\begin{figure}
\plotfiddle{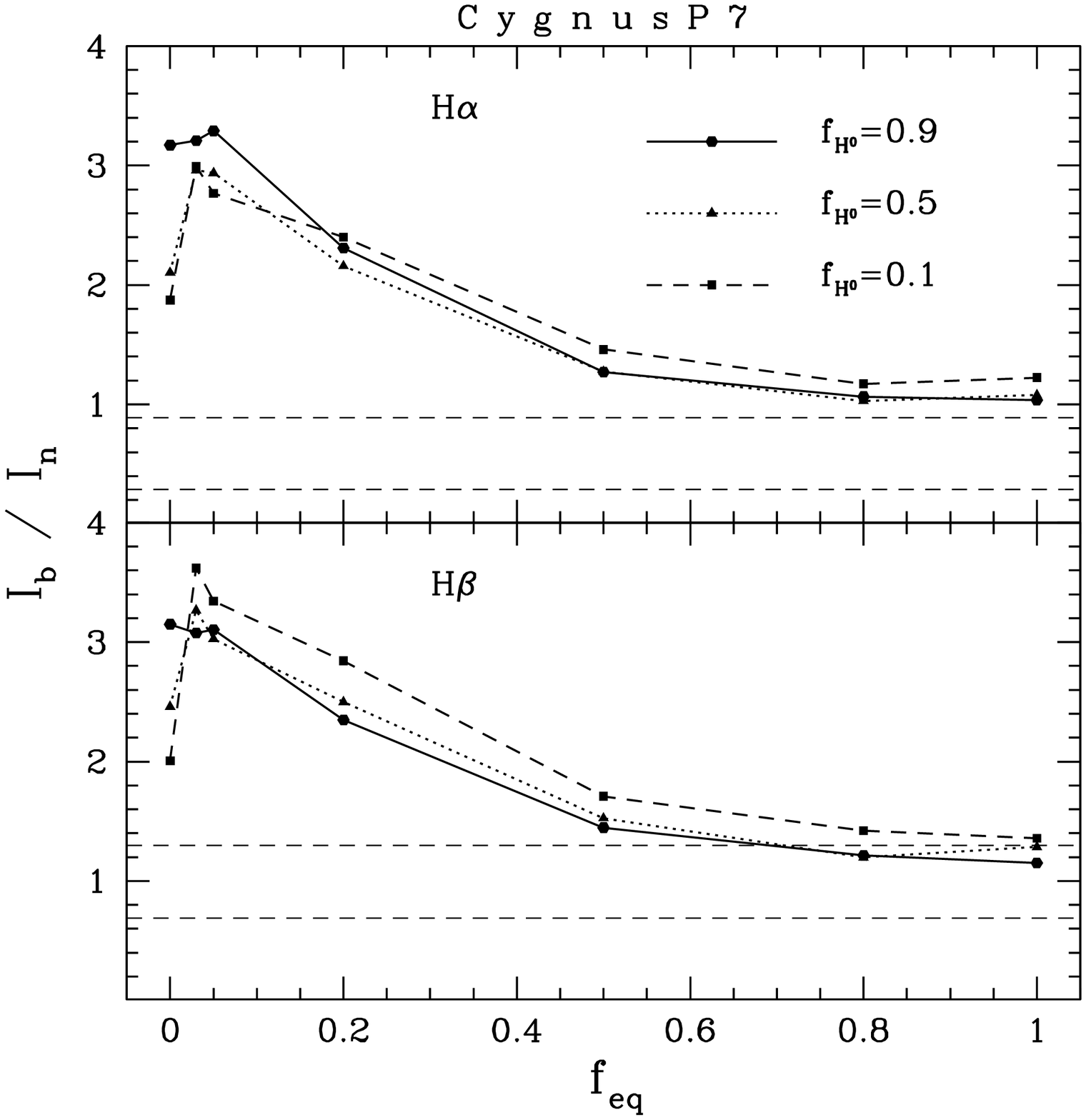}{3.in}{0.}{90.}{90.}{-280}{-240}
\end{figure}
\clearpage

\begin{figure}
\plotfiddle{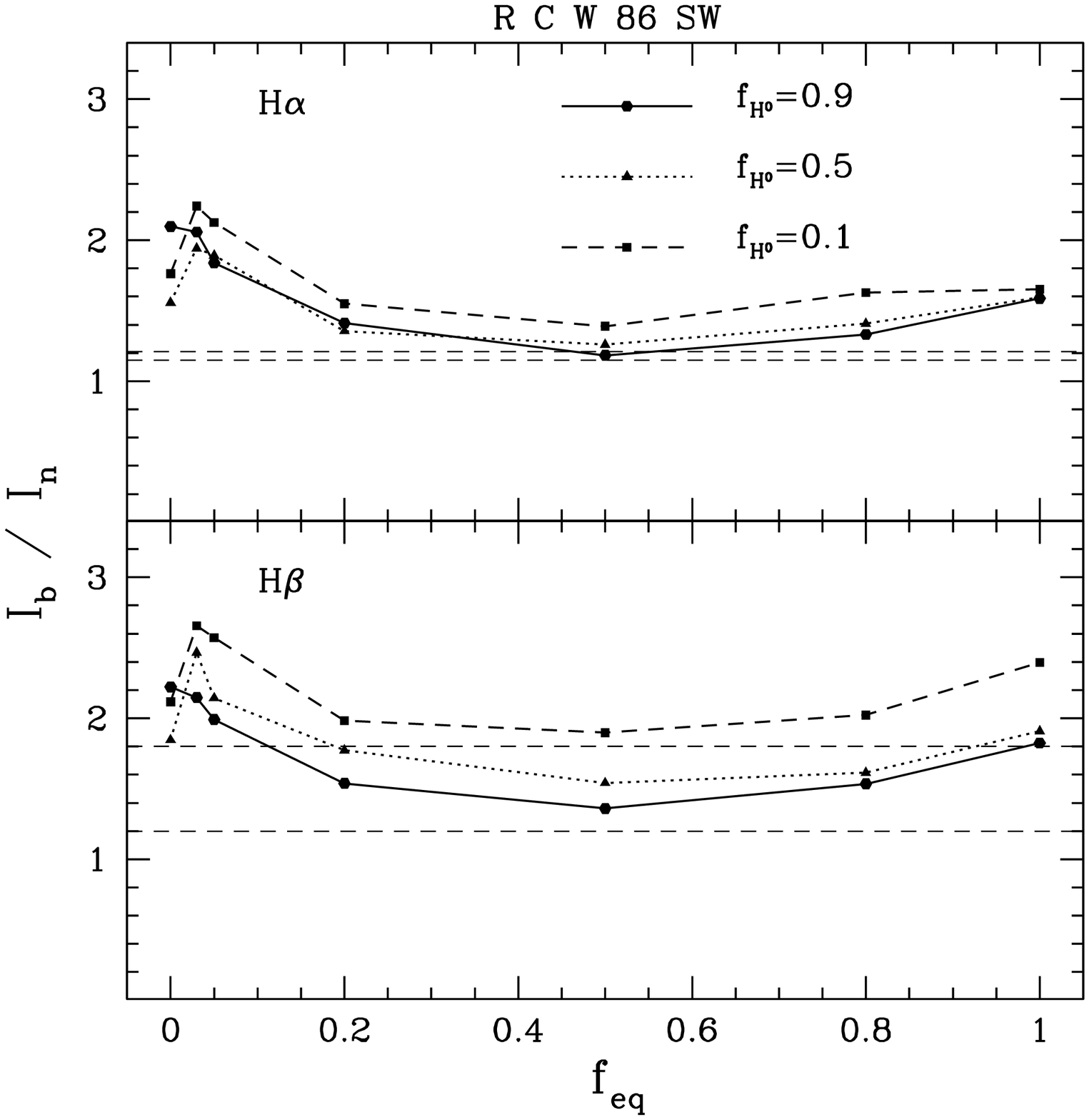}{3.in}{0.}{90.}{90.}{-280}{-240}
\end{figure}
\clearpage

\begin{figure}
\plotfiddle{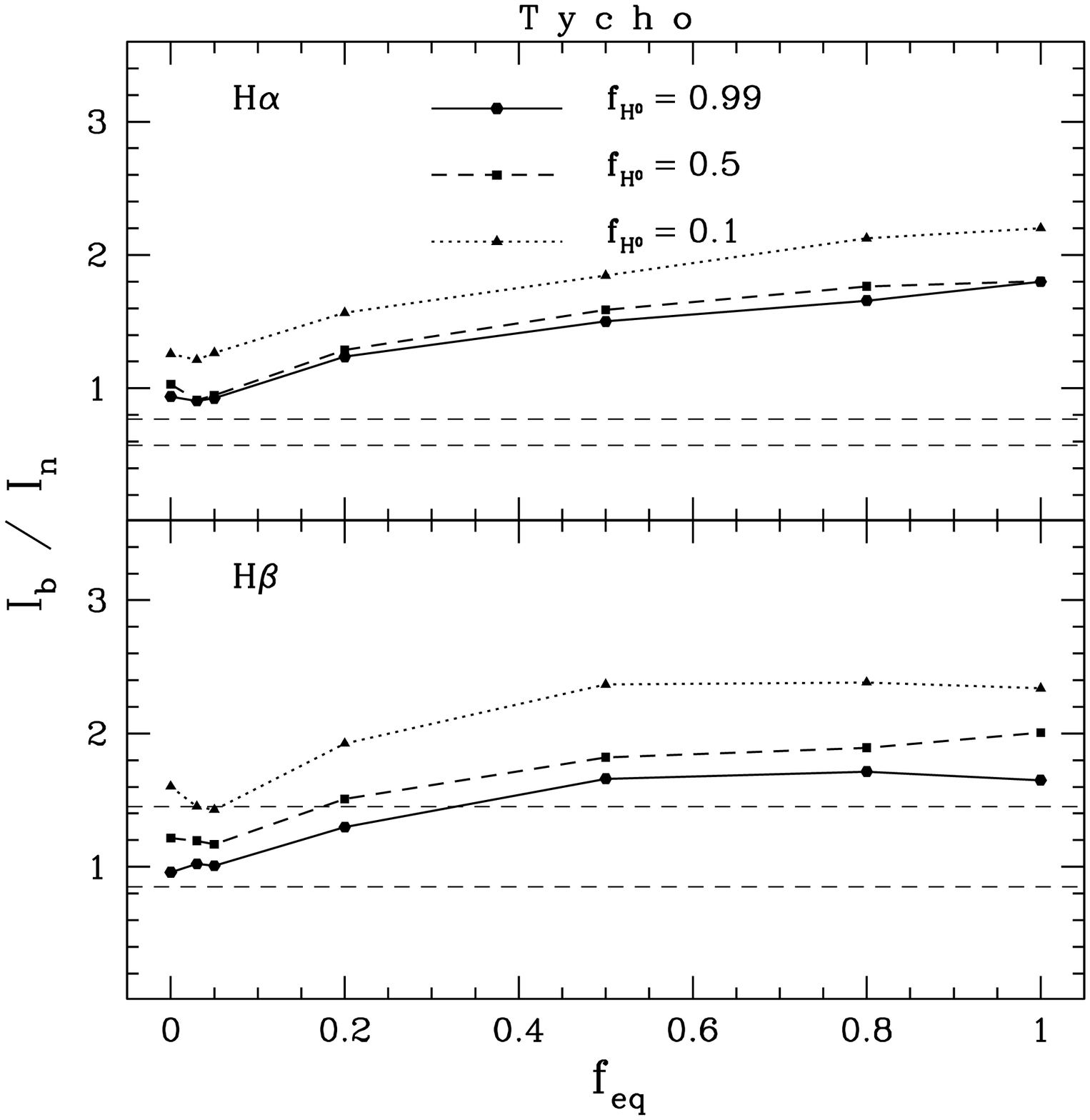}{3.in}{0.}{90.}{90.}{-280}{-240}
\end{figure}

\end{document}